\documentclass[twocolumn]{aastex631}
\begin{document}

\title{Fine structures of radio bursts from flare star AD Leo with FAST observations}

\correspondingauthor{Hui Tian}
\email{huitian@pku.edu.cn}

\author[0009-0004-0713-405X]{Jiale Zhang}
\affiliation{School of Earth and Space Sciences, Peking University, Beijing 100871, China}

\author[0000-0002-1369-1758]{Hui Tian}
\affiliation{School of Earth and Space Sciences, Peking University, Beijing 100871, China}

\author[0000-0003-1672-9878]{Philippe Zarka}
\affiliation{LESIA, Observatoire de Paris, Université PSL, Sorbonne Université, Université Paris Cité, CNRS, 92190 Meudon, France}
\affiliation{USN, Observatoire de Paris, Université PSL, Univ Orléans, CNRS, 18330 Nançay, France}

\author[0000-0002-9552-8822]{Corentin K. Louis}
\affiliation{School of Cosmic Physics, DIAS Dunsink Observatory, Dublin Institute for Advanced Studies, Dublin 15, Ireland}

\author[0000-0001-8037-5256]{Hongpeng Lu}
\affiliation{School of Earth and Space Sciences, Peking University, Beijing 100871, China}

\author[0000-0001-6643-2138]{Dongyang Gao}
\affiliation{Shandong Provincial Key Laboratory of Optical Astronomy and Solar-Terrestrial Environment, Institute of Space Sciences, Shandong University, Weihai, 264209, Shandong, China}

\author[0000-0002-3464-5128]{Xiaohui Sun}
\affiliation{School of Physics and Astronomy, Yunnan University, Kunming 650500, China}

\author[0000-0003-2872-2614]{Sijie Yu}
\affiliation{Center for Solar-Terrestrial Research, New Jersey Institute of Technology, Newark, NJ 07102, USA}

\author[0000-0002-0660-3350]{Bin Chen}
\affiliation{Center for Solar-Terrestrial Research, New Jersey Institute of Technology, Newark, NJ 07102, USA}

\author[0000-0003-2837-7136]{Xin Cheng}
\affiliation{School of Astronomy and Space Science, Nanjing University, Nanjing 210093, China}
\affiliation{Key Laboratory of Modern Astronomy and Astrophysics (Nanjing University), Ministry of Education, Nanjing 210093, China}
\affiliation{Max Planck Institute for Solar System Research, Göttingen 37077, Germany}

\author[0000-0002-7237-3856]{Ke Wang}
\affiliation{Kavli Institute for Astronomy and Astrophysics, Peking University, Beijing 100871, China}

%% Note that the \and command from previous versions of AASTeX is now
%% depreciated in this version as it is no longer necessary. AASTeX 
%% automatically takes care of all commas and "and"s between authors names.

%% AASTeX 6.31 has the new \collaboration and \nocollaboration commands to
%% provide the collaboration status of a group of authors. These commands 
%% can be used either before or after the list of corresponding authors. The
%% argument for \collaboration is the collaboration identifier. Authors are
%% encouraged to surround collaboration identifiers with ()s. The 
%% \nocollaboration command takes no argument and exists to indicate that
%% the nearby authors are not part of surrounding collaborations.

%% Mark off the abstract in the ``abstract'' environment. 
\begin{abstract}
Radio bursts from nearby active M-dwarfs have been frequently reported and extensively studied in solar or planetary paradigms. Whereas, their sub-structures or fine structures remain rarely explored despite their potential significance in diagnosing the plasma and magnetic field properties of the star. Such studies in the past have been limited by the sensitivity of radio telescopes. Here we report the inspiring results from the high time-resolution observations of a known flare star AD Leo with the Five-hundred-meter Aperture Spherical radio Telescope (FAST). We detected many radio bursts in the two days of observations with fine structures in the form of numerous millisecond-scale sub-bursts. Sub-bursts on the first day display stripe-like shapes with nearly uniform frequency drift rates, which are possibly stellar analogs to Jovian S-bursts. Sub-bursts on the second day, however, reveal a different blob-like shape with random occurrence patterns and are akin to solar radio spikes. The new observational results suggest that the intense emission from AD Leo is driven by electron cyclotron maser instability which may be related to stellar flares or interactions with a planetary companion.

\end{abstract}

%% Keywords should appear after the \end{abstract} command. 
%% The AAS Journals now uses Unified Astronomy Thesaurus concepts:
%% https://astrothesaurus.org
%% You will be asked to selected these concepts during the submission process
%% but this old "keyword" functionality is maintained in case authors want
%% to include these concepts in their preprints.

%% From the front matter, we move on to the body of the paper.
%% Sections are demarcated by \section and \subsection, respectively.
%% Observe the use of the LaTeX \label
%% command after the \subsection to give a symbolic KEY to the
%% subsection for cross-referencing in a \ref command.
%% You can use LaTeX's \ref and \label commands to keep track of
%% cross-references to sections, equations, tables, and figures.
%% That way, if you change the order of any elements, LaTeX will
%% automatically renumber them.
%%
%% We recommend that authors also use the natbib \citep
%% and \citet commands to identify citations.  The citations are
%% tied to the reference list via symbolic KEYs. The KEY corresponds
%% to the KEY in the \bibitem in the reference list below. 

\section{Introduction}\label{sec:intro}

Nearby M-type stars are among the sources of radio transients in the galaxy. The radio emission enhancement, known as the stellar radio flares or radio bursts, can exhibit very diverse characteristics, with duration from seconds to hours and frequency from megahertz ($\geqslant$10 MHz from the ground) to gigahertz. The commonly observed ones are strongly circularly-polarized radio bursts with extremely high brightness temperature. Many of them display certain structures (for instance frequency drift, e.g., \cite{2019ApJ...871..214V}) in the dynamic spectra as well. 

The emission mechanisms of stellar radio bursts have been discussed for decades. The prevailing view is that a coherent process is involved, as indicated by the brightness temperature and polarization properties. This leads to two possible mechanisms: plasma emission and electron cyclotron maser (ECM) emission \citep{1990ApJ...353..265B,2006ApJ...637.1016O,2008ApJ...674.1078O,2019ApJ...871..214V,2019MNRAS.488..559Z,2020NatAs...4..577V,2020ApJ...905...23Z,2021NatAs...5.1233C,2022ApJ...935...99B}. Both mechanisms require a certain type of unstable electron distribution and an efficient wave-particle interaction process to function \citep{1985ARA&A..23..169D,2017RvMPP...1....5M,2021MNRAS.500.3898V}. As many M-dwarfs possess strong magnetic field \citep[kilogauss photospheric magnetic field strength,][]{2021A&ARv..29....1K}, ECM emission is preferred when the cyclotron frequency exceeds the plasma frequency in the stellar coronae. Furthermore, it is commonly believed that plasma emission exhibits an o-mode circular polarization, while ECM emission is mostly characterized by x-mode circular polarization. This difference in polarization can help distinguish between the two types of emission, especially when considering the dominant magnetic polarity \citep{2019ApJ...871..214V,2021NatAs...5.1233C}.

There have been many hypotheses for the specific physical origins of coherent radio bursts, which can be broadly divided into two types, the solar paradigm and the planetary paradigm. The former proposes that the enhanced radio emission from stars is triggered by the magnetic activity, similar to the process that produces solar radio bursts, but under a different plasma environment. Routine solar monitoring has collected a variety of radio bursts and classified them into tracers of different solar energetic events like solar flares, energetic particles, and coronal mass ejections (CME). Therefore, the solar classification scheme can act as a reference for stellar radio observations. For instance, type II radio bursts and moving type IV radio bursts are most likely radio signatures of solar CMEs and may be used to trace stellar CMEs \citep{2016ApJ...830...24C,2018ApJ...856...39C,2019ApJ...871..214V,2020ApJ...895...47A,2020ApJ...905...23Z}. Though different methods have been applied to hunt for stellar CME candidates in a wide range of wavelengths \citep{2011A&A...536A..62L,2014MNRAS.443..898L,2016A&A...590A..11V,2019A&A...623A..49V,2019NatAs...3..742A,2020A&A...637A..13M,2021NatAs...5..697V,2022NatAs...6..241N,2022ApJ...933...92C,2022A&A...663A.140L,2022ApJ...928..180W}, radio detection may stand as the most hopeful smoking gun \citep{2022arXiv221105506N} in the search. However, few studies have confirmed possible solar-like radio bursts on M-dwarfs based on their morphology \citep{2020ApJ...905...23Z}. Moreover, some recent coordinated multi-wavelength observations reported radio bursts that are not time-related with flares observed in the optical band \citep{2022MNRAS.513.3482A,2022MNRAS.510.1083D}, casting doubts about their assumed connections. Another opinion is that some of the radio bursts are not produced by stellar flares, but rather result from planet-like magnetospheric dynamics of the stellar environment. Some M-dwarfs with strong global magnetic field are expected to host a magnetospheric system. They could generate radio emission in the polar region through co-rotation breakdown and star-planet interactions (SPI). The former is related to auroral magnetic-field-aligned currents due to the fast-rotating magnetosphere \citep{2012ApJ...760...59N,2017MNRAS.470.4274T} and the latter is the result of the sub-Alfvénic interactions between the star and orbiting planets \citep{2007P&SS...55..598Z,2018haex.bookE..22Z,2013A&A...552A.119S,2018ApJ...854...72T,2020NatAs...4..577V}. These types of emission are often referred to as the radio aurora. A recent study of \cite{2021NatAs...5.1233C} showed that coherent low-frequency ($\lesssim200$ MHz) radio luminosity of nearby M-dwarfs is not relevant to their level of activity, supporting the magnetospheric scenario. Co-rotation breakdown produces periodic radio signals at stellar rotation period while SPI signals are modulated by the orbital period of the planetary companion. Hence, both mechanisms require validation from the periodicity analysis through long-term radio monitoring.

Clues to the origins of stellar radio bursts may also lie in the fine structures in the dynamic spectra, which are barely understood from previous observations. In solar radio observations, coherent radio bursts usually display certain fine structures, including zebra-pattern \citep{2012ApJ...744..166T,2014ApJ...780..129T}, fiber-pattern \citep{2017ApJ...848...77W,2021A&A...653A..38W}, spike bursts \citep{1986SoPh..104...99B}, and so on (see \cite{2011ASSL..375.....C} for a review). For Jupiter, there are S-bursts \citep{1996GeoRL..23..125Z,2014A&A...568A..53R}, L-bursts \citep{1978Ap&SS..56..503R}, and zebra-pattern \citep{2018A&A...610A..69P}. It is believed that the time-resolved fine structures may provide important information about the radio source and act as diagnostic tools for the emission mechanism and electron acceleration process \citep{2007P&SS...55...89H}. However, as the radio emission from distant stars is considerably weak, the integration time is typically minutes or hours to acquire a signal of high fidelity, preventing identification of structures at second or sub-second scales. To our knowledge, only a few observations conducted by the Arecibo telescope \citep{1989A&A...220L...5G,1997A&A...321..841A,2006ApJ...637.1016O,2008ApJ...674.1078O} and Effelsberg radio telescope \citep{2001A&A...374.1072S} are capable of revealing the fine structures of stellar radio bursts in detail. Above all, \cite{2006ApJ...637.1016O,2008ApJ...674.1078O} presented the clearest and finest radio emission structures to date which are characterized by the presence of many short-duration, fast-drifting sub-bursts. Since then, there have been no significant updates in this field.

The Five-hundred-meter Aperture Spherical radio Telescope \citep[FAST,][]{2006ScChG..49..129N,2011IJMPD..20..989N} is the largest single-dish radio telescope in the world, which aims to detect the weak signals in the universe with extraordinary sensitivity. In this research, we present the results from the first high time-resolution observations on an M-dwarf AD Leo with FAST. The paper is structured as follows. In section \ref{sec:data}, we introduce our observations and data reduction procedures. In Sections \ref{sec:obs1} and \ref{sec:obs2}, we present the observational results on Dec. 2nd and Dec. 3rd, 2021, respectively. Section \ref{sec:int} discusses the emission mechanism of the radio bursts and gives some possible interpretations of the origin. We further discuss our results in Section \ref{sec:dis} and summarize the research in Section \ref{sec:sum}. 

\section{Observations and data reduction}\label{sec:data}
The FAST observations were granted in the approved observation project PT2021\_0019. We focus on the two rounds of 3-hour FAST observations which were conducted on Dec. 2nd and Dec. 3rd in 2021, 20:30 UT to 23:30 UT at each night. We observed the flux calibrator 3C286 10 minutes after each observation of AD Leo. We also coordinated optical observations with Xinglong 85 cm telescope and Weihai 1 m telescope on Dec. 3rd.

\subsection{Target star: AD Leo}
AD Leo is an M3.5V star with a mass of 0.42\(M_\odot\) \citep{2008MNRAS.390..567M} and radius of 0.44\(R_\odot\) \citep{2015ApJ...804...64M}, at a distance of 4.965 pc from the solar system \citep{2020yCat.1350....0G}. Its rotation period is $2.230\pm0.001$ days \citep{2023arXiv230203377F}. AD Leo is an extensively studied flare star whose flaring activities have been detected in radio \citep{1986ApJ...305..363L,1989A&A...220L...5G,1990ApJ...353..265B,1997A&A...321..841A,2001A&A...374.1072S,2006ApJ...637.1016O,2008ApJ...674.1078O,2019ApJ...871..214V}, visible light \citep{1991ApJ...378..725H,2003ApJ...597..535H,2006A&A...452..987C,2020A&A...637A..13M}, extreme ultraviolet \citep{1997ApJ...491..910C,2003ApJ...582..423G}, and X-ray bands \citep{1999A&A...342..502S,2003A&A...411..587V,2005A&A...435.1073R}. AD Leo has a quiet radio emission of about 2mJy \citep{1989A&A...210..284J} and can grow almost 500 times brighter during flaring time. Previous Zeemann-Doppler imaging (ZDI) studies have suggested that AD Leo has a predominant dipole magnetic component and a nearly pole-on geometry, with the magnetic south pole visible over a long time \citep{2008MNRAS.390..567M,2018MNRAS.479.4836L}. Recent ZDI measurements show that AD Leo maintains a 70\% dipolar component with some contributions of higher order terms (Bellotti et al., in prep, personal communication with Julien Morin).

Recently, there has been great interest in determining whether AD Leo hosts a potential exoplanet. Based on the analysis of radial velocity (RV) variation, \cite{2018AJ....155..192T} claimed the possible existence of a hot giant planet with a mass of $0.237 \pm 0.047\; M_{\rm jup}$ in spin-orbit resonance (period of 2.23 days). However, the validity of this claim has been questioned by some subsequent studies \citep{2020A&A...638A...5C,2022A&A...666A.143K}, which ascribed the RV modulation to stellar activities.

\subsection{Radio observations from FAST}

FAST is a 500 m radio telescope located in a karst depression in Guizhou province, China, with an illuminated aperture of 300 m \citep{2019SCPMA..6259502J}. The 19-beam L-band receiver \citep{2020RAA....20...64J} is currently in commission for FAST which works at frequencies from 1000 MHz to 1500 MHz with full polarization measurements from two linear feeds. We chose the pulsar backend mode with a sampling time of 196.608 $\mu s$ and 1024 frequency channels for our observations. We tracked our target star with 'tracking with angle' observation mode which focused the target with the central beam and compensated the rotation of the sky view during the observation. A series of noise diode signals with a known temperature of $\sim$12 K were injected for $\sim$1 s every $\sim$16 s, which is intended for flux and polarization calibration. The data from other beams was also collected.

\subsection{FAST data reduction pipeline}\label{sec:fdr}

The FAST data reduction pipeline includes several key procedures which are:

(1) Noise subtraction: We calculated the injection time of the noise diode signals and subtracted them using the difference of the reading value with the noise 'on' and adjacent 'off' states.

(2) Polarization and flux calibration: We used the noise diode signals to correct the mismatches between the gains and phases of the two linear feeds. The amplitude of the signals was compared to the reported noise temperature to determine the conversion coefficient between the digital output and the antenna temperature (K). We used the calibrator's observations to derive the absolute gain (the conversion coefficient between the antenna temperature (K) and the flux density (Jy)). More details are provided in Appendix \ref{sec:pol}.

(3) Radio frequency interference (RFI) flagging: We masked the frequency channels that are severely affected by RFIs. These channels have a prominently larger standard deviation in the radio flux density and are not shown in the observation parts. We provided two clean frequency bands (a group of frequency channels within a certain frequency range) which are 1004-1146 MHz and 1293-1464 MHz, though some channels in the two bands may still be occasionally disrupted by narrow-band RFIs.

(4) Background removal: Both Stokes \emph{I} and Stokes \emph{V} radio data contain background contributions that slowly vary throughout the 3-hour observation period. In addition, Stokes \emph{I} data experience another broadband, rapidly-changing contribution likely due to the fluctuating system temperature. To reveal the real emission variations over short time intervals ($1-2$ minutes), we performed a linear fit of the flux density variation within these intervals for each frequency channel and subtracted it from both Stokes \emph{I} and Stokes \emph{V} data. Next, we used a quiet frequency band, with no radio bursts or strong RFIs, as a background for a second stage of subtraction, but only for the Stokes \emph{I} data. Specifically, we defined a 50 MHz wide frequency band with a minimum average flux density as the quiet band. The first stage of background subtraction may potentially remove some slowly-varying radio emission from the target star, while the second stage of background subtraction may remove some broadband ($>$500 MHz) variability in the flux density. Both subtractions have little impact on the fine structures discussed in this paper.

\subsection{Optical observations with Xinglong 85 cm telescope and Weihai 1 m telescope}

The photometric observation on Dec. 3rd was conducted by the 85 cm telescope \citep{2009RAA.....9..349Z,2018RAA....18..107B} at Xinglong Station of National Astronomical Observatories, Chinese Academy of Sciences. The telescope has been equipped with a Johnson-Cousins $UBVRI$ filter system and a 2048$\times$2048 pixel CCD camera with a $0.93''$ pixel scale since the last upgrading \citep{2018RAA....18..107B}. The observation lasted for $\sim$ 4.5 hours, from 18:13 UT to 22:47 UT. The CCD images were processed with the standard procedures in the IRAF package \citep{1999ASPC..189...35D}, including bias subtraction, flat-field correction, cosmic ray removal, and aperture photometry. We produced the light curves of the star in the form of relative magnitude in B, V, R bands.

We applied the Weihai 1 m telescope \citep{2014RAA....14..719H,2016PASP..128l5002G} which is located at Weihai Observatory of Shandong University, for spectroscopic observation on the same day. It is designed with a Cassegrain optical system and mounted with a fiber-fed high-resolution Echelle spectrograph \citep{2016PASP..128l5002G}. The spectrograph covers the spectral region 371 -- 1100 nm and has a spectral resolution of 40,600 -- 57,000. The observation started from 17:57 UT to 21:30 UT and the exposure time for each spectrum is 30 minutes. The raw spectroscopic data were reduced using standard procedures in IRAF packages \citep{1999ASPC..189...35D}, including bias subtraction, flat fielding, spectra extraction, scattered light subtraction, wavelength calibration, continuum normalization, and heliocentric velocity correction.

\section{Observations on Dec. 2nd, 2021}\label{sec:obs1}

\begin{figure*}
	\centering
	\includegraphics[width=\linewidth]{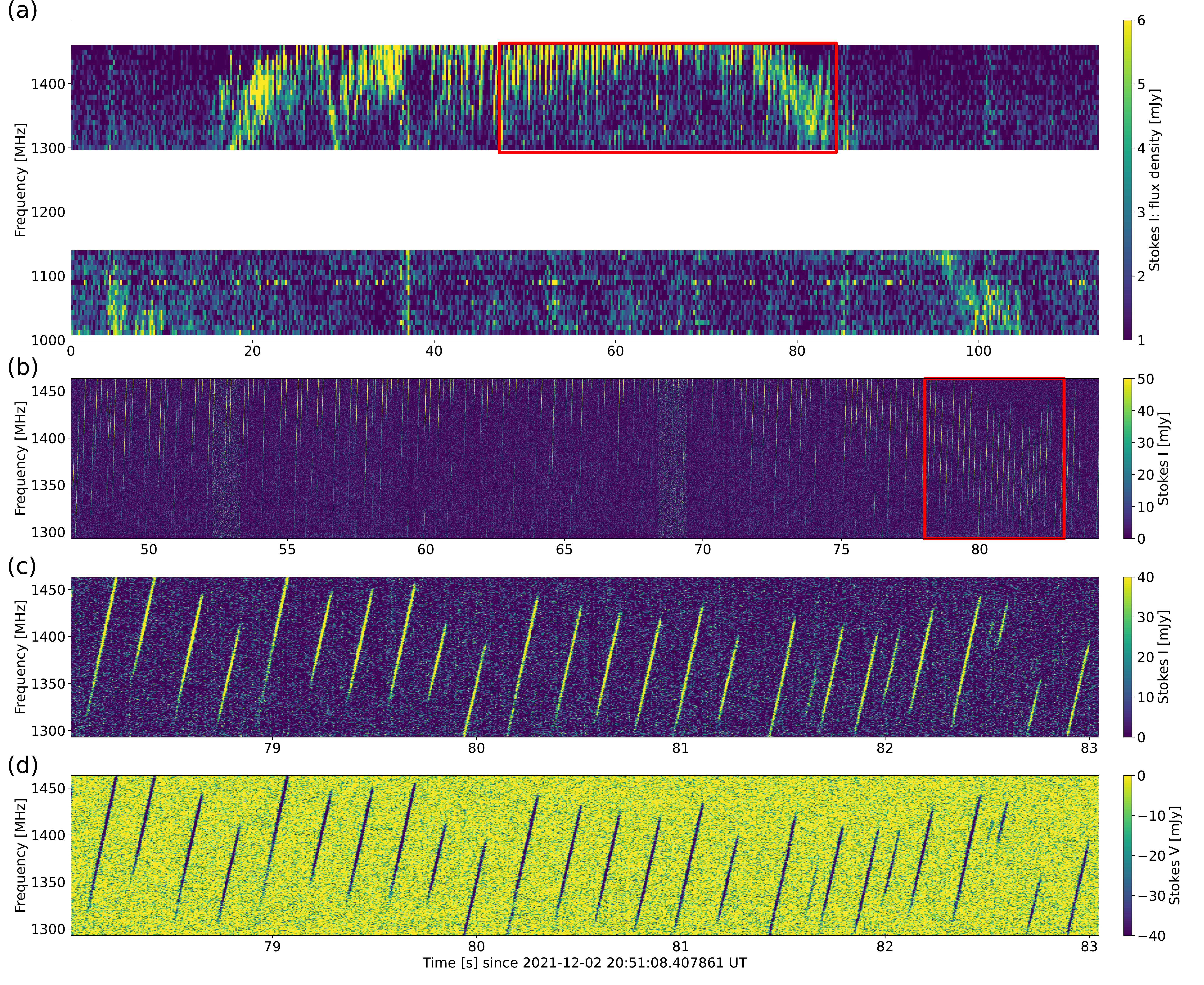}
	\caption{Dynamic spectra of a radio burst observed on Dec. 2nd. (a) Low-resolution Stokes \emph{I} dynamic spectrum of the event. The data was rebinned by a factor of 1024 in time and 16 in frequency which gives the time resolution of 0.2 s and frequency resolution of 7.8 MHz. The red rectangle represents the time and frequency range of the dynamic spectrum in panel (b). Gaps are due to the RFI excision. (b) High-resolution Stokes \emph{I} dynamic spectrum. The data was rebinned by a factor of 32 in time, giving a time resolution of 6.3 ms. The frequency dimension was not rebinned, giving a resolution of 0.49 MHz. The one-second long vertical diffuse features are due to the increased background fluctuation when the noise diode signals are injected. The red rectangle represents the time range for panel (c). (c) Details of the Stokes \emph{I} dynamic spectrum. The time and frequency resolutions are the same as in panel b. (d) Corresponding Stokes \emph{V} dynamic spectrum. The time and frequency resolutions are also the same as in panel (b). Positive value indicates left-hand circular polarization and negative value indicates right-hand circular polarization.\label{fig:figure1}}
\end{figure*}

FAST detected multiple intense radio bursts at the beginning (20:45 UT -- 20:53 UT) of the 3-hour observation with no detection made at later time. The strongest event was observed at around 20:51 UT, as is illustrated in Figure \ref{fig:figure1}. A low-resolution dynamic spectrum is presented in panel (a) with a time resolution of $\sim$0.2 s and frequency resolution of $\sim$7.8 MHz. The radio emission lasts for about 100 s and mainly occurs at frequencies of 1300 -- 1450 MHz, though it starts and ends at lower frequencies ($\sim$1000 MHz). A positive frequency drift around 0 -- 25 s and a negative one around 70 -- 100 s can be noticed despite the data gap. The drift rates were estimated as about 28 MHz/s and -19 MHz/s, respectively. We estimated the noise fluctuation level using the standard deviation of the radio flux density during quiet time. The noise level is $\sim$0.9 mJy and the measured maximum signal is $\sim$15 mJy, giving a maximum signal-to-noise ratio (SNR) at $\sim$ 17. Detailed sub-structures can be seen in the dynamic spectrum, but they are not fully resolved at this resolution. 

We used a shorter data rebinning to generate a dynamic spectrum with a higher resolution. As is shown in panel (b) in Figure \ref{fig:figure1}, the dynamic spectrum has a time bin size of $\sim$ 6 ms and a frequency bin size of $\sim$ 0.49 MHz. We can see that the event contains many spiky radio sub-bursts at a very short time scale. Their morphology is further shown in panel (c). The radio sub-bursts share a stripe-like shape and very consistent drift from lower frequencies to higher frequencies. They come in different sizes, and most of them last for about 0.1 s and cover 100 MHz. Furthermore, the occurrence of the sub-bursts shows $\sim$0.2 s quasi-periodicity in the selected time range. The strongest radio signal reaches an intensity of 188 mJy, which yields an SNR of 10 with a noise level of 18 mJy at this resolution. It is interesting to note that the SNR does not decrease much when the time and frequency resolutions are improved by at least one order of magnitude. The reason is that the radio emission is emitted in a very short period and the second-long integration time will eventually smooth the data and underestimate the real intensity.

We adopted the polarization convention commonly used in pulsar astronomy \citep{2010PASA...27..104V} which defines Stokes \emph{V} as left-hand circularly polarized light minus right-hand circularly polarized light. As shown by the negative value in the Stokes \emph{V} dynamic spectrum in panel (d), the radio sub-bursts are all right-hand circularly polarized, some of which can reach a polarization degree of nearly 100\%. Radio bursts observed on this day are all structured in the form of uniformly drifting sub-bursts shown in Figure \ref{fig:figure1}, therefore we do not elaborate on other events here.

\begin{figure*}
	\centering
	\includegraphics[width=0.8\linewidth]{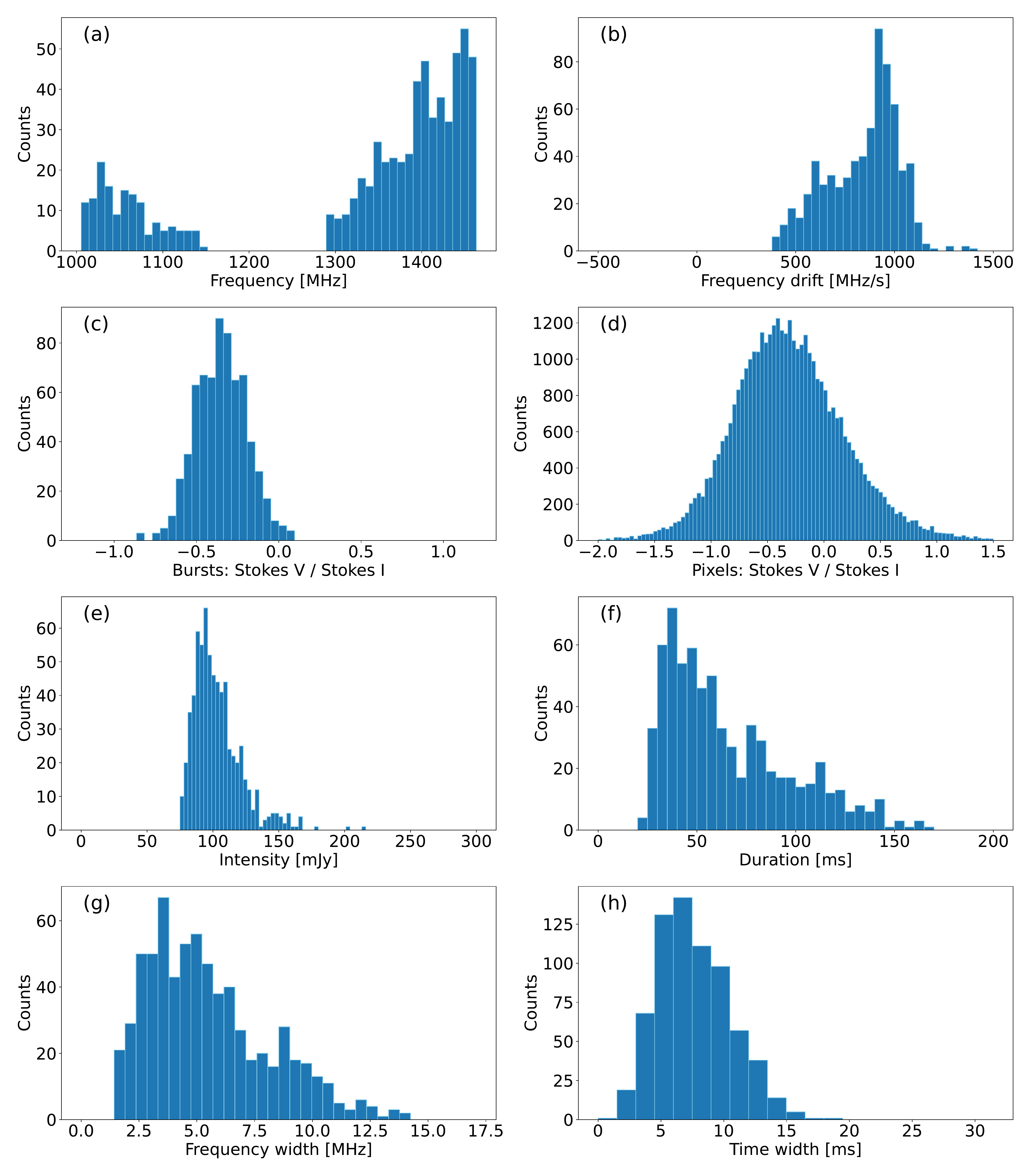}
	\caption{Histograms of different properties of the radio sub-bursts observed on Dec. 2nd. (a) Central frequency. (b) Frequency drift rate. (c) The ratio between Stokes \emph{V} and Stokes \emph{I} average intensity of the sub-bursts. (d) The ratio between Stokes \emph{V} and Stokes \emph{I} intensity for each pixel in the sub-bursts. The unphysical values ($|V/I|>1$) are mainly due to the measurement error. (e) The average intensity of the sub-bursts. (f) Duration. Duration defined as the total lasting time when the intensity exceeds the 3$\sigma$ threshold. (g) Frequency width. Frequency width defined using the number of channels over the 3$\sigma$ threshold at peak time. (h) Time width. Time width defined using the number of time steps over the 3$\sigma$ threshold at central frequency. The methods to compute these parameters are further explained in Appendix \ref{sec:pro}.\label{fig:figure2}}
\end{figure*}

To statistically analyze the properties and behaviors of the enormous radio sub-bursts, we developed a method to automatically detect all the radio sub-bursts in the dynamic spectra on this day. The methodology is explained in Appendix \ref{sec:pro}. The distributions of the sub-bursts are shown in Figure \ref{fig:figure2} according to their properties like central frequency, frequency drift rate, degree of circular polarization, intensity, duration, frequency width, and time width. Frequency width refers to the instantaneous bandwidth of the sub-burst at the time of its peak, time width refers to the fixed-frequency duration (at central frequency) of the sub-burst and duration refers to its overall lifetime. Potential sub-bursts occurring between 1146 MHz and 1293 MHz were not considered. The drift rate peaks around 900 MHz/s and the degree of circular polarization peaks around 35\%. Most of the radio sub-bursts have an average intensity of around 100 mJy and a duration of 0.02 -- 0.15 s. The peak of the time (frequency) width is 6 ms (3.5 MHz).

\begin{figure*}
	\centering
	\includegraphics[width=0.6\linewidth]{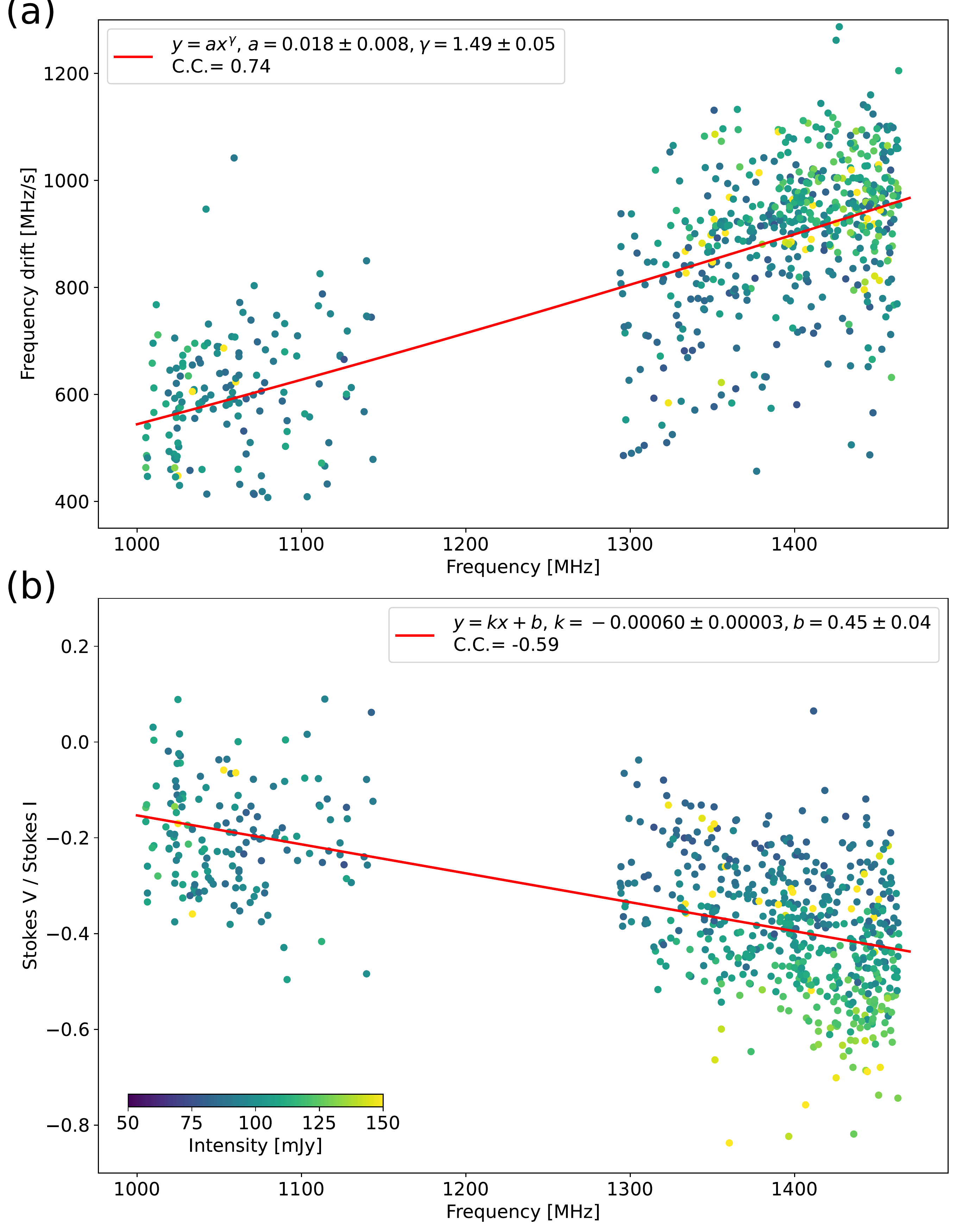}
	\caption{Scatterplots of the Dec. 2nd radio sub-bursts in the parameter space. (a) Scatterplot of the frequency drift of the sub-bursts against their frequency. The color represents the intensity of the sub-bursts. The best fit result and the correlation coefficient (C.C.) are included in the legend. (b) Scatterplot of the ratio between the Stokes \emph{V} and Stokes \emph{I} intensity of the sub-bursts against their frequency. \label{fig:figure3}}
\end{figure*}

Despite this, we found that the drift rate and circular polarization degree of the sub-bursts are dependent on their frequency, which is illustrated by the scatterplots in Figure \ref{fig:figure3}. Sub-bursts with a higher frequency tend to display a larger frequency drift rate. A trend is also seen in the bottom panel, where a higher frequency implies a larger degree of circular polarization. We use a power-law expression and a linear expression, respectively, to quantitatively characterize the two relationships. The former gives a power law index of $1.49\pm 0.05$ and the latter gives a slope of $-0.00060\pm0.00003$. The correlation coefficients are 0.74 and -0.59, respectively.

\begin{figure*}
	\centering
	\includegraphics[width=0.8\linewidth]{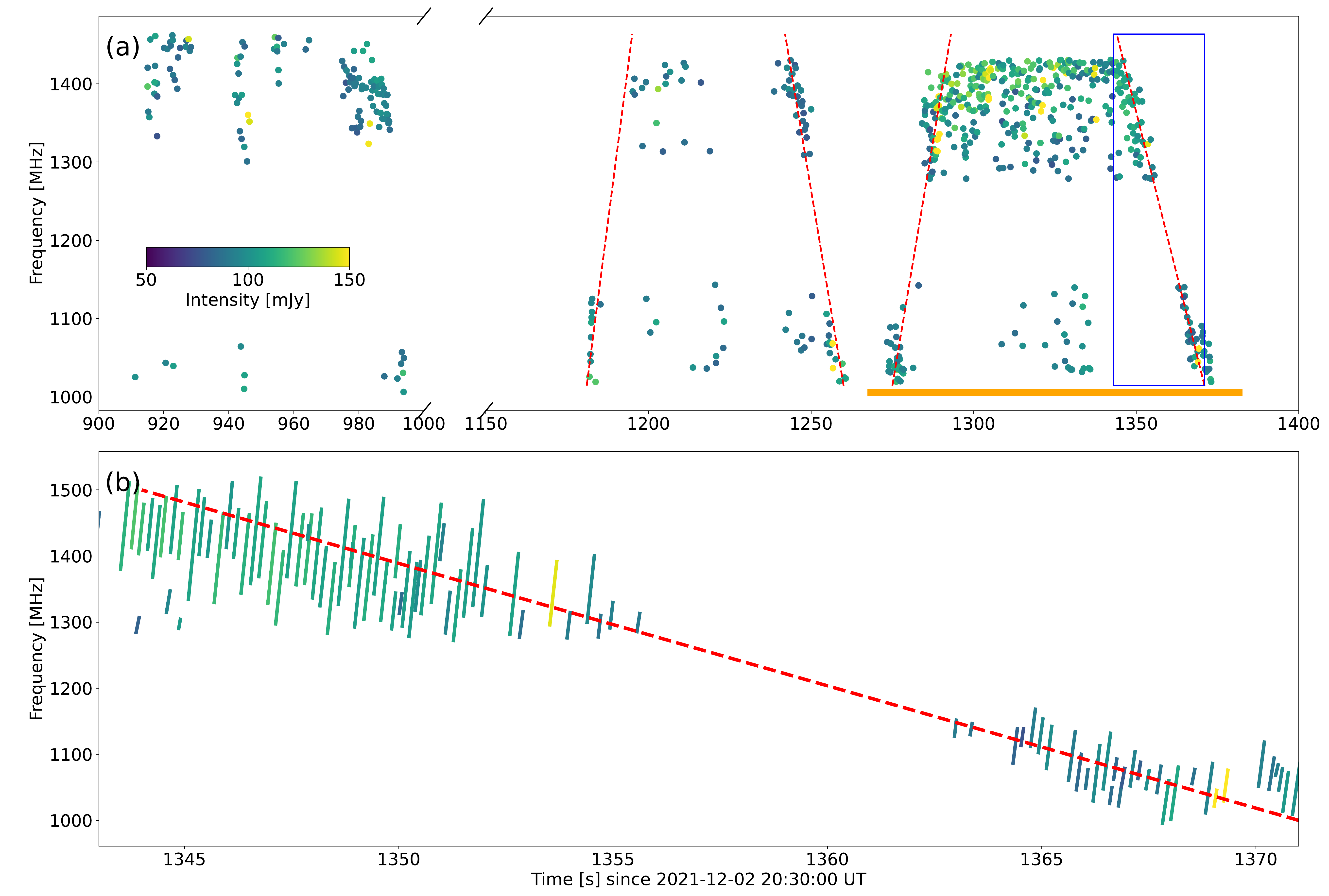}
	\caption{Distribution of the radio sub-bursts observed on Dec. 2nd in the time-frequency domain. (a) Scatterplot of the radio sub-bursts. The color represents the intensity of the sub-bursts. The dashed lines indicate the four slow-drifting radio sub-burst trains. The marked frequency drift rates are 36 MHz/s, -28 MHz/s, 28 MHz/s, and -19 MHz/s, respectively. The orange bar represents the time range of panel (a) in Figure \ref{fig:figure1}. (b) Details of a radio sub-burst train in the blue rectangle region. The line segments represent the simplified morphology of the radio sub-bursts. \label{fig:figure4}}
\end{figure*}

We also checked the distribution of radio sub-bursts in the time-frequency domain to see if there is any overall pattern in their occurrence. Figure \ref{fig:figure4}(a) shows all the locations of the identified sub-bursts during the observation period. The most prominent feature is the trains of sub-bursts whose trends are marked with dashed lines. Each train contains many sub-bursts that drift to higher frequencies at a constant rate, as is demonstrated in Figure \ref{fig:figure4}(b). As a whole, their central frequencies reveal a smaller drift rate at 19 -- 36 MHz/s. We refer to them as radio sub-burst trains. There are four of them and one with positive frequency drift appears at first and the negative one follows. Therefore, we suppose that the radio sub-burst trains may actually come in pairs with a positive component and a negative one. The two radio sub-burst trains probably connect at higher frequencies and make up an inverted V-shape. However, the joint points are blocked by the bandpass limit of our observation. If they exist, the joint frequencies deduced from the trends of the sub-burst trains give 2234 MHz and 2067 MHz, which might stand for the upper frequency limit of the emission.

\section{Observations on Dec. 3rd, 2021}\label{sec:obs2}

\begin{figure*}
	\centering
	\includegraphics[width=\linewidth]{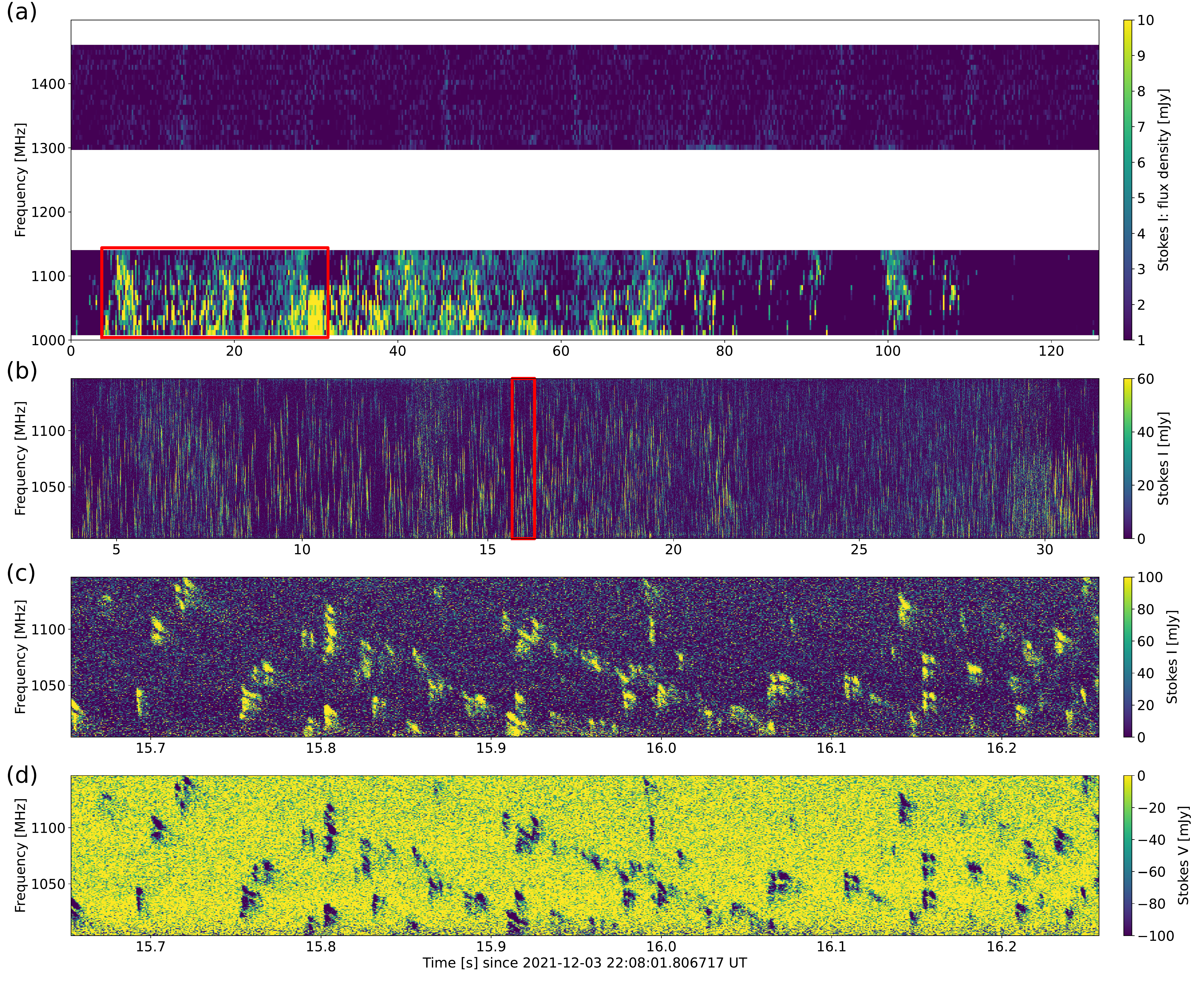}
	\caption{Dynamic spectra of a radio burst observed on Dec. 3rd. (a) Low-resolution Stokes \emph{I} dynamic spectrum of the event. The data was rebinned by a factor of 1024 in time and 16 in frequency which gives the time resolution of 0.2 s and frequency resolution of 7.8 MHz. The red rectangle represents the time and frequency range of the dynamic spectrum in panel (b). Gaps are due to the RFI excision. (b) High-resolution Stokes \emph{I} dynamic spectrum. The data was rebinned by a factor of 32 in time, giving a time resolution of 6.3 ms. The frequency dimension was not rebinned, giving a resolution of 0.49 MHz. The red rectangle represents the time range for panel (c). (c) Details of the Stokes \emph{I} dynamic spectrum. The data was rebinned by a factor of 4 in time, giving a time resolution of 0.8 ms. The frequency resolution is 0.49 MHz. (d) Corresponding Stokes \emph{V} dynamic spectrum. The time and frequency resolutions are the same as in panel (c).\label{fig:figure5}}
\end{figure*}

In the Dec. 3rd observation, the detected radio emission lasted for about 1.5 h (21:13 UT -- 22:48 UT) in the 3-hour observation period and showed up as many distinct radio bursts with different timescales. In Figure \ref{fig:figure5}, we present the dynamic spectra of the strongest radio event on this day. As is shown in the top panel, the radio emission mainly comes from the lower frequency band of 1000 -- 1150 MHz. The highest emission intensity is 36 mJy and the noise level is 0.8 mJy, giving an SNR of 45. In a close-up image in panel (b), a great number of spiky sub-bursts can be noticed. They have smaller sizes and occur much more frequently compared with Dec. 2nd observation. Fine structures in panel (c) reveal that each distinguishable radio sub-burst has a blob-like shape unlike the previous ones. Many of them exhibit a slight elongation in the time-frequency domain, with a typical duration of about a few milliseconds and a majority of drift patterns from higher to lower frequencies. Some sub-bursts are regularly lined up and form a radio sub-burst train with a negative overall frequency drift (for instance from 15.8 s to 15.9 s) while others occur in a seemingly random manner. More particularly, we notice that the sub-bursts tend to gather in pairs or clusters. These radio sub-burst pairs have one component in higher frequencies and another one in lower frequencies with a separation of only a few MHz wide. The strongest radio burst can reach an intensity of 680 mJy. The noise level at the time resolution of $\sim$0.8 ms is 45 mJy which results in an SNR of 15. Same with the observation on Dec. 2nd, the emission on Dec. 3rd also displays a high degree of right-hand circular polarization, which is evident in the bottom panel.

\begin{figure*}
	\centering
	\includegraphics[width=\linewidth]{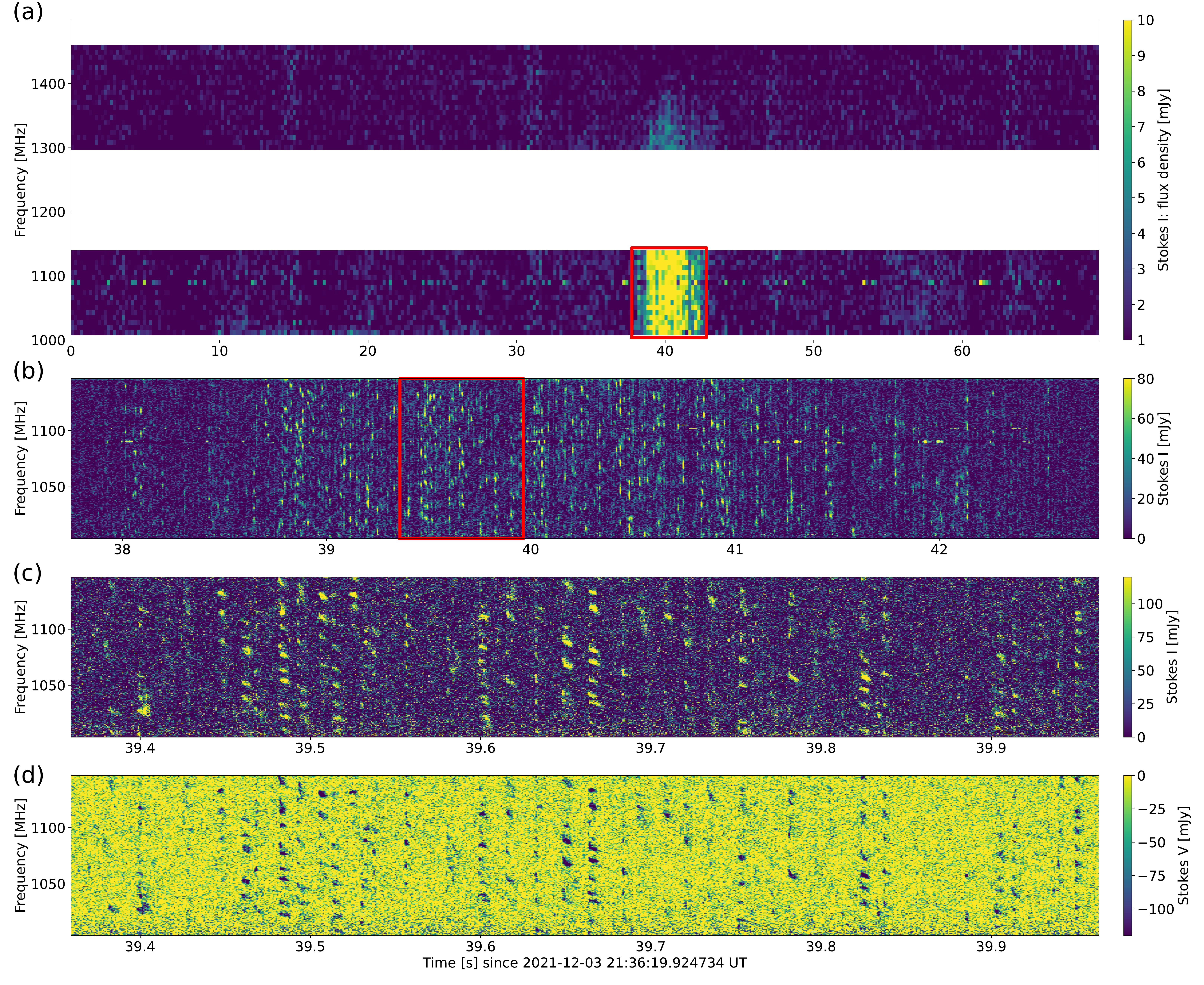}
	\caption{Dynamic spectra of another radio burst observed on Dec. 3rd. The settings in each panel are the same as in Figure \ref{fig:figure5}. The red rectangle in panel (a) represents the time and frequency range of panel (b) and the one in panel (b) represents the time range of panel (c) and (d).\label{fig:figure6}}
\end{figure*}

The radio burst in Figure \ref{fig:figure5} is a representative of the events observed on this day. Almost all the events show enhanced emission below 1150 MHz and structure in the form of blob-like sub-bursts. We are not intended to go through all the cases, but we want to mention another event in particular, which has radio emission extending above 1300 MHz and displays a strange gathering pattern of the sub-bursts. As is shown in Figure \ref{fig:figure6}, the radio event only lasts for 5 s. The individual radio sub-bursts have a blob-like shape as well, but they do not form any frequency-drifting trains as mentioned before. Instead, many of the sub-bursts emerge at the same time and approximately equally-spaced frequencies, forming many vertical stripes in the dynamic spectrum. This example demonstrates the diversity and complexity of the sub-bursts' behaviors in the time-frequency domain.

\begin{figure*}
	\centering
	\includegraphics[width=0.8\linewidth]{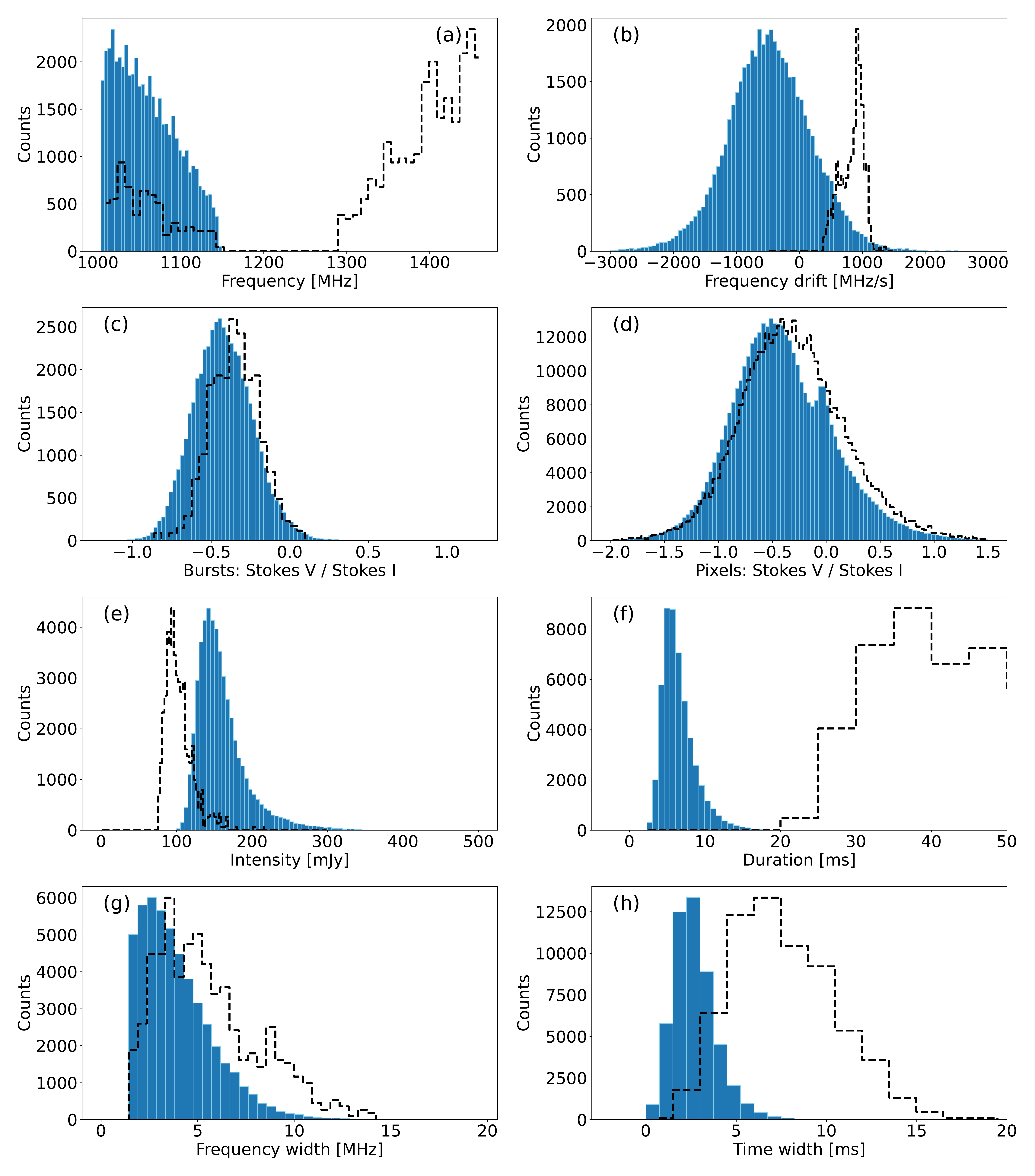}
	\caption{Histograms of different properties of the radio sub-bursts observed on Dec. 3rd. The parameters in different panels are the same as in Figure \ref{fig:figure2}. The shapes of the histograms in Figure \ref{fig:figure2} are shown as black dashed lines for comparison, after scaling by a factor to match the peak amplitude in the plot.\label{fig:figure7}}
\end{figure*}

Radio sub-bursts were automatically searched and statistically analyzed under the same methodology mentioned in Section \ref{sec:obs1}. Their property distributions are shown in Figure \ref{fig:figure7}. Most of the sub-bursts are detected at 1000 -- 1150 MHz. A negative frequency drift rate is in the majority with a peak at $\sim$ -450 MHz/s. The degree of circular polarization is centered around 45\% and the intensity is mostly around 140 mJy. The typical duration is 2 -- 15 ms. The distribution peak of the time (frequency) width is 2.5 ms (2.5 MHz). The properties of the radio sub-bursts observed on these two days are summarized in Table \ref{tab:table1}. We attempted to find some correlations between the parameters, but failed to reach any reliable results similar to those in Figure \ref{fig:figure3}.

\begin{table*}
    \centering
    \caption{Properties of the radio sub-bursts observed on two days.\label{tab:table1}}
    \begin{tabular*}{0.7\textwidth}{@{\extracolsep{\fill}}lcc}
    \hline
    Date & Dec. 2nd & Dec. 3rd \\
    \hline
    \hline
    Occurrence time &20:45 -- 20:53 UT &21:13 -- 22:48 UT\\
    \hline
    Frequency range (MHz) & 1000 -- 1500 & 1000 -- 1150(mostly) \\
    \hline
    Frequency drift rate (MHz/s) & $\sim$900 (positive) & $\sim$450 (negative) \\
    \hline
    Degree of circular polarization (\%) &$\sim$35 &$\sim$45\\
    \hline
    Intensity (mJy) &$\sim$100 &$\sim$140\\
    \hline
    Duration (ms) &20 -- 150 &2 -- 15\\
    \hline
    Frequency width (MHz) &$\sim$3.5 &$\sim$2.5\\
    \hline
    Time width (ms) &$\sim$6 &$\sim$2.5\\
    \hline
    \end{tabular*}
\end{table*}

\begin{figure*}
	\centering
	\includegraphics[width=0.8\linewidth]{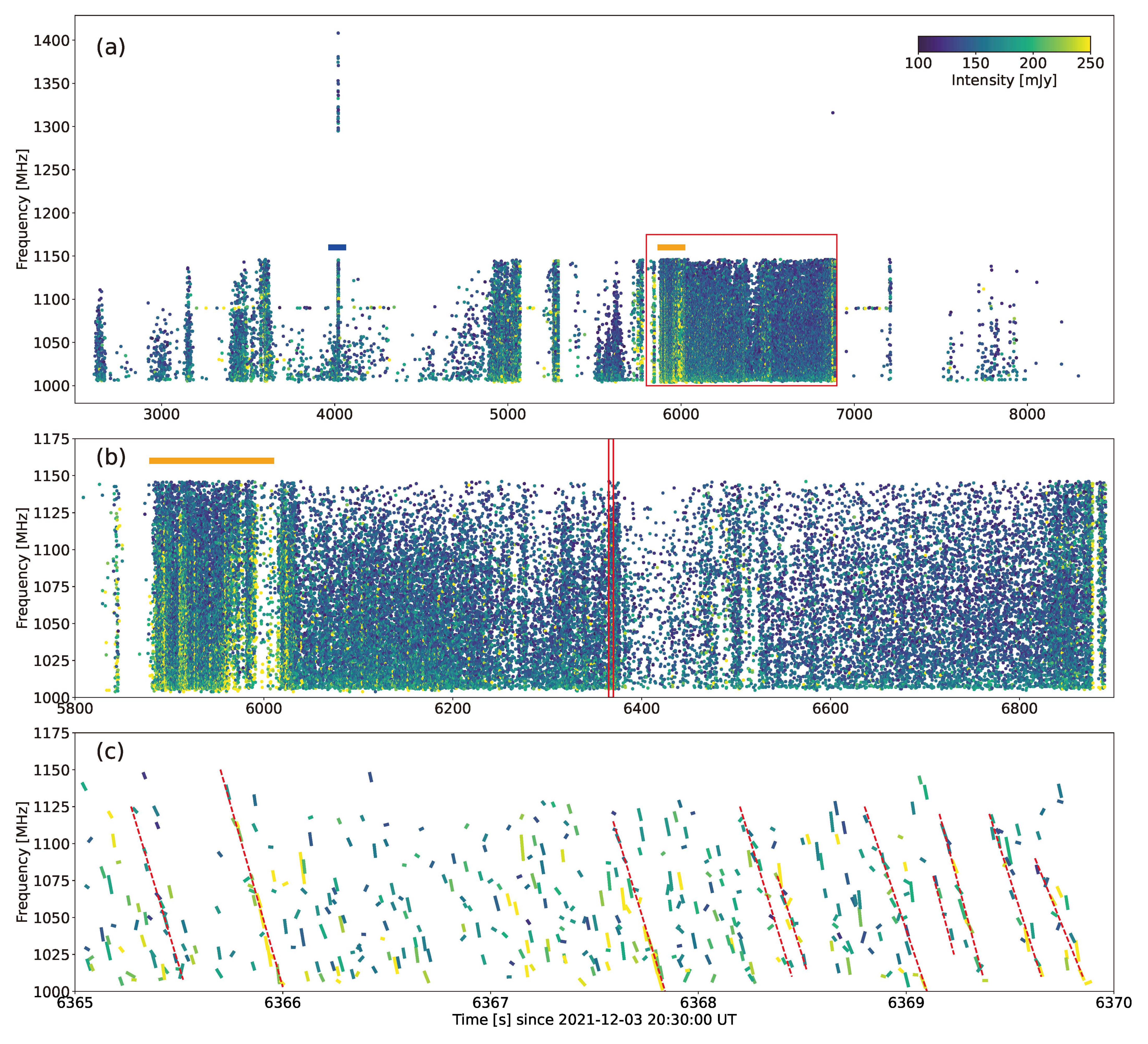}
	\caption{Distribution of the radio sub-bursts observed on Dec. 3rd. in the time-frequency domain. (a) Scatterplot of the radio sub-bursts. The orange (blue) bar represents the time range of panel (a) in Figure \ref{fig:figure5} (Figure \ref{fig:figure6}). The red rectangle marks the time and frequency range of a long-duration radio burst. (b) Details of the long-duration radio event. The orange bar represents the time range of panel (a) in Figure \ref{fig:figure5}. The red rectangle here indicates a time range which is further demonstrated in panel (c). (c) Close-up morphology and distribution of the radio sub-bursts in the given time range. Several radio sub-burst trains are marked by the dashed lines.\label{fig:figure8}}
\end{figure*}

We then turned to analyze the distribution of the radio sub-bursts in the time-frequency domain. As is shown in panel (a) in Figure \ref{fig:figure8}, we found several isolated radio pulses, each of them has a timescale from seconds to minutes and consists of many millisecond-scale sub-bursts. Among them, the longest and the strongest event marked by the red rectangle lasts for $\sim$ 18 minutes, part of which has been shown in Figure \ref{fig:figure5}. A close-up look in panel (b) reveals many pulsation structures with denser dots in the image. An example of a radio pulse is shown in panel (c) with more details. Many sub-bursts seem to follow the drifting pattern while others are irregular. The overall drift rates of the sub-burst trains are around -500 MHz/s, close to the value of the individual sub-bursts.

\begin{figure*}
	\centering
	\includegraphics[width=0.8\linewidth]{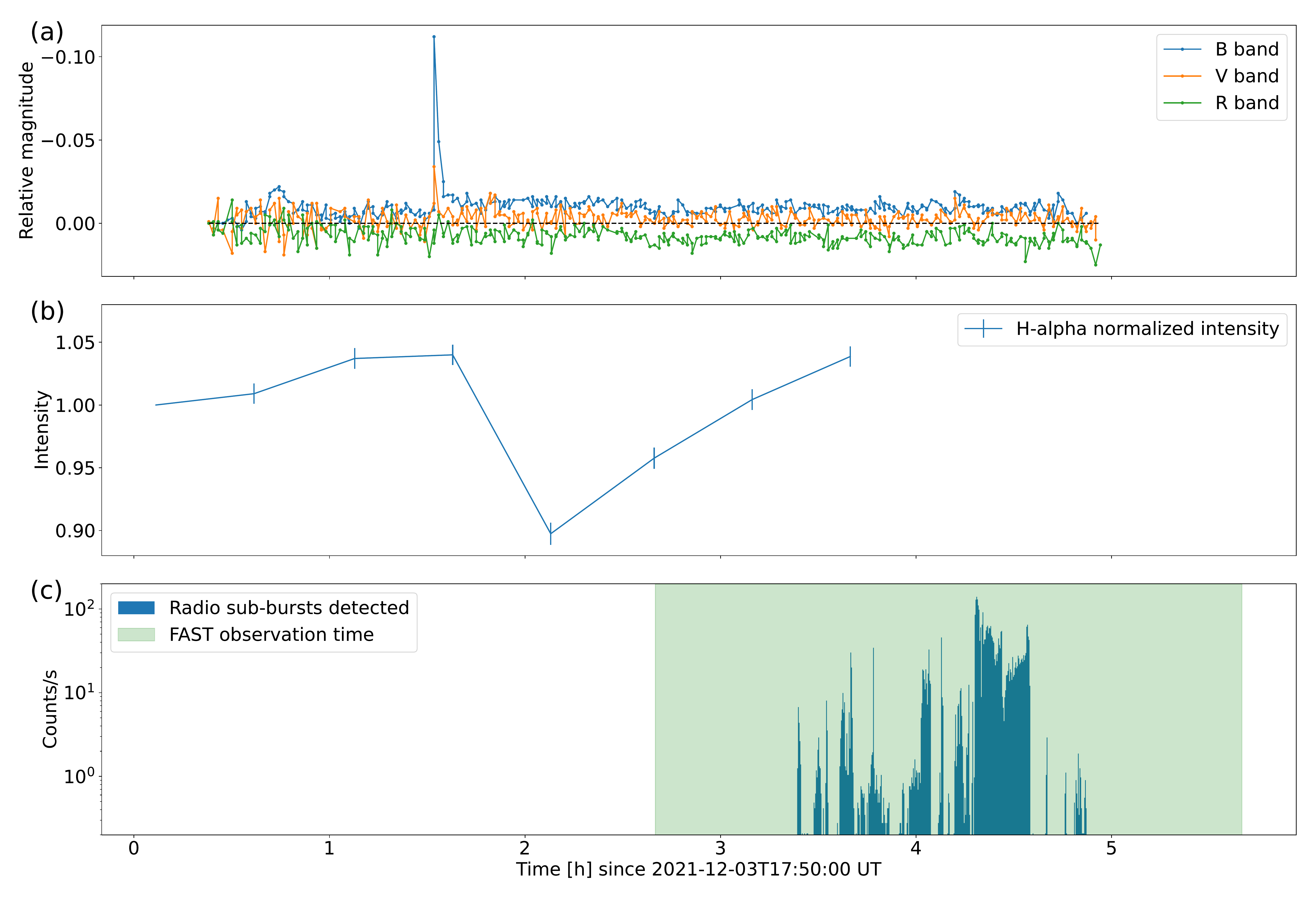}
	\caption{Comparison of the radio and optical observations on Dec. 3rd. (a) Light curves in B, V, and R bands from the photometric observation of the Xinglong 85 cm telescope. The magnitude was compared with the value at the beginning of each band. The dashed black line indicates the zero level. (b) H-alpha normalized intensity from the spectroscopic observation of the Weihai 1 m telescope. The intensity was obtained by integrating the spectrum of 6561.68 -- 6563.75 \AA. The error was derived from the standard deviation of the continuous spectrum. (c) Detected number of radio sub-bursts per second in the FAST observation. The shaded region indicates the time range of the available FAST observation.\label{fig:figure9}}
\end{figure*}

The results of the photometric and the spectroscopic observations in the optical-band are presented in Figure \ref{fig:figure9}. In panel (a), a flare signal is detected around 19:22 UT in B-band and V-band. H-alpha line intensity from the spectroscopic observation also displays some variations (a mild increase and a sharp decrease) at the time around. However, we have no radio information on the stellar flare as it is not covered by the FAST observation. The detected radio bursts happened 2 -- 3 hours after the event, during which no clear flaring activities can be identified from the light curves. Therefore, our multi-wavelength observation suggests no evident correlation between the optical flare and the radio flare.

\section{Interpretations}\label{sec:int}

\subsection{Emission mechanism}\label{sec:emimec}

The detected radio emission has an intensity of order 100 mJy. Assuming that the radio source has a spatial upper limit equivalent to the stellar disk, the lower limit of the brightness temperature is estimated to be on the order of $10^{11}$ K. The circular polarization of the radio emission also suggests that it is coherent in nature. Hence, our discussion will be limited to two types of emission: plasma emission and ECM emission. Plasma emission is produced by the electron beam injecting into the background plasma and is usually seen in solar type II and type III radio bursts. Some high-resolution observations of type II and type III radio bursts have identified some fragmented structures like the second-long, narrow-band striae or spikes \citep{2017NatCo...8.1515K,2020ApJ...897L..15M,2021NatAs...5..796R,2022URSL....4...17B}. The millisecond-scale sub-bursts in our observations have a much shorter duration and are probably not related to the sub-structures of plasma emission. ECM emission, on the other hand, is likely responsible for some more short-lived sub-bursts (down to a few milliseconds), with narrow frequency bandwidths and recurring behaviors. A typical example is the Jovian S-bursts consisting of many uniformly frequency-drifting emission segments with an instantaneous duration of a few milliseconds and a total duration of tens or hundreds of milliseconds \citep{1996GeoRL..23..125Z,2001P&SS...49..365Q,2014A&A...568A..53R}. Interestingly, the observed sub-bursts on Dec. 2nd closely resemble the Jovian S-bursts in terms of discreteness and uniformity of the frequency drift rate, despite the fact that S-bursts usually exhibit negative frequency drifts. The morphology and distribution of the blob-like sub-bursts on Dec. 3rd are similar to solar radio spikes which feature many small-scale, impulsive millisecond spikes randomly occurring in the time-frequency plane \citep{1986SoPh..104...99B,2007ApJ...665L.171W}. While there is still controversy surrounding the emission mechanism of solar radio spikes, ECM mechanism has more advantages in explaining their properties and has been explored in numerous theoretical modelings \citep{1994SSRv...68..159M,1998PhyU...41.1157F}.

\cite{2008ApJ...674.1078O} reported many radio spectral structures of AD Leo with bandwidths and duration similar to those in our observations. Based on their calculations, they argued that if the plasma emission is relevant, the bandwidth and duration require a relatively cool and dense plasma that is neither efficient enough to produce the intense emission nor transparent enough to let the emission escape. Additionally, \cite{2021MNRAS.500.3898V} compared the typical bandwidth and duration of the two types of coherent emission in the stellar coronae. He believed that the rise-time of the plasma emission is largely subject to the conversion rate from Langmuir waves to transverse electromagnetic waves, which will prevent generating radio bursts shorter than a second. In contrast, the ECM mechanism is predicted to generate emission in a very short time scale and narrow bandwidth in many studies. For instance, \cite{2006A&ARv..13..229T} attributes the narrow bandwidth to microscopic elementary radio sources and the short time scale to the extraordinarily high maser growth rate. The spectral and temporal properties may be intrinsic to the microscopic and transient events producing the ECM emission. Therefore, while both plasma emission and ECM emission can produce highly structured patterns in the dynamic spectrum, the ECM mechanism is much more effective in amplifying radiation in sub-second or millisecond scales, making it a more likely explanation for the observed fine structures.

To further investigate the application of the plasma emission and ECM emission at different coronal heights, we derived the radial profiles of the two types of emission frequency, which only depend on the plasma density and the magnetic field strength. We adopted the measured density at $\log{n_e}=10.4$ based on the O~{\sc{vii}} line ratios (most sensitive to $\sim$ 3 MK plasma) from the X-ray observations \citep{2004A&A...427..667N} as the base density of the corona and obtained the evolution with height using the constant-gravity hydrostatic equilibrium (HSE) model following the frameworks described in \cite{2017ApJ...845...67C} and \cite{2019ApJ...871..214V}. We adopted the coronal temperature range of 2-10 MK from \cite{2004ApJ...613..548M} and chose three different temperatures (3 MK, 6 MK, 9 MK) for our discussion. The density follows an exponential decrease with distance and the scale height is given by $H={k_BT}/{(\mu m_H g)}$, where $k_B$ is the Boltzmann constant, $\mu$ is the mean molecular weight which we adopted as 0.6, $m_H$ is the mass of the hydrogen atom and $g$ is the gravitational acceleration at the stellar surface. We derived $H=0.23r_0, 0.45r_0, 0.68r_0$ ($r_0$ as the stellar radius) for coronal temperatures of 3 MK, 6 MK, 9 MK, respectively. The magnetic field profile was modeled by assuming that the strength follows a cubic decay with distance, $B=B_0 r^{-3}$. We referred to the results from the recent ZDI modeling (Bellotti et al., in prep) where a purely dipole fit to the Stokes V profiles gives a maximum amplitude of $923\pm 70$ G at the magnetic south pole. We chose $B_0=920$ G in the polar surface, $B_0=460$ in the equatorial surface (half of the polar magnetic field strength in a dipole case), and the value in between to constrain the analysis.

\begin{figure*}
	\centering
	\includegraphics[width=0.6\linewidth]{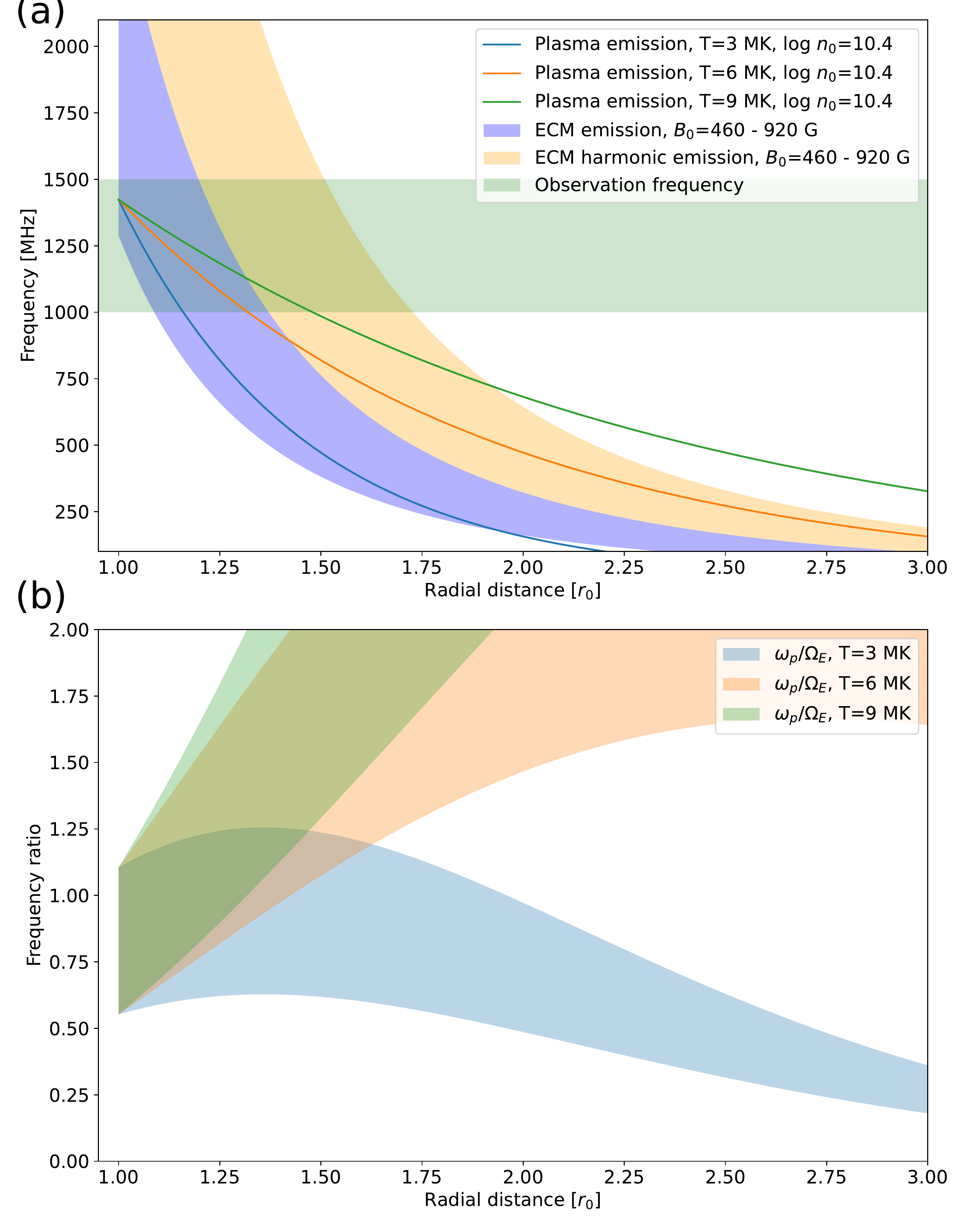}
	\caption{Radial profiles of the plasma emission frequency and the ECM emission frequency. The radial distance is the distance to the stellar center in the unit of the stellar radius ($r_0$). (a) Solid lines with different colors indicate the frequencies of the plasma emission for different coronal temperatures. Blue and orange regions indicate the frequency range of ECM fundamental and the second harmonic emission, respectively. The green region denotes the frequency range of the FAST L-band observation. $B_0$ denotes the magnetic field strength at the stellar surface and $T$ denotes the coronal temperature. (b) Range of the ratio between the plasma frequency and cyclotron frequency for different coronal temperatures. \label{fig:figure10}}
\end{figure*}

The production of the plasma emission requires the local plasma frequency ($\omega_p$) to exceed the electron cyclotron frequency ($\Omega_E$), and vice versa for the ECM emission. Figure \ref{fig:figure10} displays the frequency of the two types of coherent emission as a function of radial distance. From panel (a), the accountable source of the 1000 -- 1500 MHz emission is predicted to be 1 -- 1.5 $r_0$ to the core (0 -- 0.5 $r_0$ above the surface) for both plasma emission and ECM fundamental emission. In this height range, the plasma frequency is comparable to the cyclotron frequency. Near the equator, the plasma frequency is slightly larger than the cyclotron frequency, while in the polar region, it is slightly smaller. In this case, the ECM emission should only be supported at high latitudes where a larger $B_0$ can meet the $\omega_p/\Omega_E<1$ requirement. However, we presume that there might be some low-density regions similar to the auroral plasma cavities which allow a much lower $\omega_p/\Omega_E$ value. As supported by the auroral kilometric radiation (AKR) observations, the electric potential drops in these regions can deplete background plasma and create favorable condition for ECM instability \citep{1979GeoRL...6..479B,1979ApJ...230..621W,2006A&ARv..13..229T}. It is also possible that small-scale magnetic field structures exist and raise the local magnetic strength. Our model also does not consider the complexities introduced by the stellar wind and centrifugal force. Besides, we cannot fully rule out the possibility of the ECM harmonic emission which could operate at $\omega_p/\Omega_E>1$ plasma \citep{1984JGR....89..897M,1985JGR....90.9663W}. Harmonic emission may occur at higher altitudes in the corona, around 1.5 $r_0$ to the core for the second harmonic emission indicated by Figure \ref{fig:figure10}(a). To sum up, the analysis on the fine structures and modeling on the coronal environment favor ECM emission as the mechanism behind the radio bursts.

\subsection{Phenomenon explanations}\label{sec:pheno}
To facilitate the following analysis, we assumed the ECM fundamental emission as the mechanism for the radio emission.

\subsubsection{Sub-burst drift rate}
The radio sub-bursts have a typical frequency drift rate of several hundreds of MHz/s. The most likely explanation is the source motion. As the frequency only depends on the magnetic field strength ($f=eB/(2\pi m_e)\approx 2.80 B$ [MHz]), the frequency drift could be related to the magnetic field gradient which is shown in the following equation.

\begin{equation}
    \frac{d f}{d t}=2.80\frac{d B}{d t}=2.80 v \frac{d B}{d s}.
\end{equation}

$v$ is the source motion velocity and $s$ is the length of the trajectory. We assumed that the magnetic field follows $B=B_0(r/r_0)^{-3}$ and the velocity is purely in the radial direction. At a height of 0.2 stellar radius from the stellar surface and magnetic field strength of 446 G (corresponding frequency of 1250 MHz), the magnetic field gradient in the radial direction is estimated as 0.0036 G/km. Then, the frequency drift rate at $\sim$900 MHz/s on the first day yields a velocity of 0.3 $c$ and $\sim$450 MHz/s on the second day yields 0.15 $c$, which could be transformed to the electron energy at 25 keV and 6 keV, respectively. The positive frequency drift should correspond to the downward propagating velocity and the negative one should represent the upward motion. Note that the estimated electron energy can only be regarded as a lower limit as we fully neglect the perpendicular electron velocity.

The frequency-drift rate relationship in Figure \ref{fig:figure3}(a) can be explained under the same paradigm. Keeping the hypotheses above the same and assuming that the velocity is invariant with height, we found that the drift rate is proportional to $r^{-4}$ and the emission frequency is proportional to $r^{-3}$. Hence, the power-law relationship between the frequency and the drift rate can be easily obtained.

\begin{equation}
    \frac{d f}{d t} \propto f^{4/3}
\end{equation}

The discrepancy between the power index $4/3$ and the measured value 1.49 may arise from the oversimplification of the model. First of all, we did not consider the real trajectories of the electrons. They are supposed to gyrate along the magnetic field lines which are likely not in the radial direction. Secondly, the electrons are expected to follow the adiabatic motion in which the magnetic moment ($\mu_e=W_{\perp}/B$, $W_{\perp}$ as the perpendicular kinetic energy) is conserved. The assumption of a constant parallel velocity may fail in this regard. Thirdly, the electric potential drop might exist and influence the electron parallel velocity along the field lines \citep{2007P&SS...55...89H}. Similarly, there has also been a $f$-$df/dt$ relationship found for Jovian S-bursts and a more precise analytical model has been accepted \citep{1996GeoRL..23..125Z,1999A&A...341..918G}. Based on a known magnetic dipole field, $f$-$df/dt$ relationship is determined by the electron energy, L-shell, and electron pitch angle at the equator. A detailed investigation in this framework will be presented in a follow-up paper.

The $f$-$df/dt$ relationship observed on Dec. 2nd (Figure \ref{fig:figure3}(a)) is not re-observed in the Dec. 3rd observation. This may be related to the very different morphology of the fine structures on that day (blobs), itself possibly related to a different structure of the plasma in the sources, revealing a dynamic stellar plasma environment. But we note that drift-rate measurements of Dec. 2nd are strongly scattered in Figure \ref{fig:figure3}(a), and that the average drift rate for Dec. 3rd is consistent with that on Dec. 2nd, suggesting the same range of electron energies (that will be studied in detail in the follow-up paper). More observations are clearly required to explore the variability of stellar radio bursts.

 \subsubsection{Time and frequency widths of the sub-bursts}

The time width and bandwidth of the radio sub-bursts can be used to estimate the spatial scale of the radio source. The time width of the radio sub-burst at a certain frequency is equivalent to the passing time of the source at a certain location. Therefore, the spatial scale of the radio emitter responsible for the individual radio sub-burst can be estimated as

\begin{equation}
    \triangle r_t=v\triangle t
\end{equation}

where $\triangle t$ is the time width. The bandwidth of the radio bursts can also help to constrain the estimation.

\begin{equation}
    \triangle r_f=\frac{\triangle f}{d f/d r}=\frac{v\triangle f}{d f/d t}
\end{equation}

For the radio sub-bursts on Dec. 2nd, we took $\triangle t=6\;\rm{ms}$, $\triangle f=3.5\;\rm{MHz}$, $v=0.3\;c$, $df/dt=900\;\rm{MHz/s}$, which yields $\triangle r_t=5\times 10^{2} \;\rm{km}$, $\triangle r_f=3\times 10^{2} \;\rm{km}$. For the radio sub-bursts on Dec. 3rd, $\triangle t=2.5\;\rm{ms}$, $\triangle f=2.5\;\rm{MHz}$, $v=0.15\;c$, $df/dt=450\;\rm{MHz/s}$ yields $\triangle r_t=1\times 10^{2} \;\rm{km}$, $\triangle r_f=2\times 10^{2} \;\rm{km}$. Therefore, the radio sources on Dec. 3rd have a relatively smaller size. Compared with the reported results from planetary observations, we found that our estimation is close to the characteristic size of Jovian decametric and hectometric radio emission (DAM and HOM) \citep{1970ApJ...159..671D,2017GeoRL..44.4439L,2018GeoRL..45.9408L,2020GeoRL..4790021L}.

With the estimate of the rough size of the radio source, we can determine the magnitude of the brightness temperature. Considering an intensity of 100 mJy, a frequency of 1200 MHz, and a spatial scale of $1\times 10^{2}\;\rm{km}$, the brightness temperature is on the order of $10^{18}\;\rm{K}$.

\subsubsection{Circular polarization}
The sense of circular polarization is thought to match the magnetic field polarity of the radio source. \cite{2019ApJ...871..214V} found a dominant left-hand circular polarization in long-duration radio events of AD Leo and suggested that it is consistent with x-mode emission from the visible magnetic south pole. However, there were also reports on right-hand circularly polarized emission of AD Leo \citep{2008ApJ...674.1078O,2021NatAs...5.1233C} which means that the polarization origins can be quite diverse. In our two days of observations, the radio emission displays right-hand circular polarization, which means x-mode emission if the emission comes from the northern hemisphere and o-mode from the southern hemisphere. O-mode emission is supported by our estimation of the ratio between the plasma frequency and the cyclotron frequency at low altitudes (o-mode for $0.5<\omega_p/\Omega_E<1$, \cite{1984JGR....89..897M,1985JGR....90.9663W}) while x-mode requires relatively smaller density or higher magnetic strength. As the ECM emission is highly beamed, observability of the two hemispheres should be discussed with reference to the large-scale magnetic field geometry.

We also noted that the emission is only $\sim$ 40\% circularly polarized and the degree is dependent on the frequency (suggested by Figure \ref{fig:figure3}(b)). \cite{2021NatAs...5.1233C} reported a $\sim$ 41\% degree of circular polarization of AD Leo at 144 MHz and argued that the plasma emission is potentially possible, which we do not consider here. \cite{2022ApJ...935...99B} offered two other explanations for fractional circular polarization in stellar radio emission: two separated radio sources with opposite senses of circular polarization or depolarization effect. We believe that the former does not apply to our observations, as the polarization degree is intrinsic to individual sub-bursts in which only one source is involved. The latter, however, is possible and the prediction that the depolarization is weaker at higher frequencies (thus polarization is stronger) agrees with our observations. It is also possible that the fractional circular polarization is related to $\omega_p/\Omega_E>0.2$ plasma condition which accounts for $<100$\% x-mode or o-mode emission. The changes in polarization degree with the frequency may be related to the spatial variation of the frequency ratio shown in Figure \ref{fig:figure10}(b).

\subsubsection{Quasi-periodicity}
Previous radio observations of AD Leo have revealed many quasi-periodic pulsations with periods ranging from tens of milliseconds to several seconds \citep{1986ApJ...305..363L,1989A&A...220L...5G,1990ApJ...353..265B,2001A&A...374.1072S,2008ApJ...674.1078O}. The cause of the periodicity is not fully understood, but it may be associated with intermittent electron accelerations. On Dec. 2nd, we detected a quasi-periodic sub-burst train similar to series of repeating Jovian S-bursts, which have a frequency of 10 -- 100 Hz \citep{2014A&A...568A..53R}. \cite{2006JGRA..111.6212E} argued that those repeating S-bursts are the result of periodic electron accelerations by the inertial Alfvén waves resonating in Jupiter’s ionosphere. The statement was supported by subsequent simulation works \citep{2007JGRA..11211212H,2009GeoRL..3614101H} which could reproduce the periodic occurrence and comparable morphology of Jovian S-bursts. The quasi-periodic radio sub-burst series in our observation may leave a hint on a $\sim$ 5 Hz Alfvén waves which exert a periodic modulation to the electron population. However, it is also worth noting that at other times during the same day, we saw a lot of non-periodic isolated sub-bursts. Additionally, sub-bursts on the second day take on totally different characteristics, which suggests that multiple competing processes might be at play, resulting in diverse spectral structures.

\subsubsection{Inverted V-shape pattern}
The inverted V-shape patterns in Figure \ref{fig:figure4} have frequency drifts an order of magnitude slower than the sub-burst frequency drifts. They are probably not induced by the individual motion of the accelerated electrons. Instead, they may be related to the bulk motion of the emission region. A possible explanation is the rotational modulation. Jupiter’s observations reveal that the ECM emission is beamed to a very thin radiation cone ($\sim1^{\circ}$ ) and sweeps across the observer as Jupiter rotates similar to the lighthouse beacon. The frequency drift can be a geometric effect when the emission in different locations rotates to the line-of-sight direction over time. However, the beaming pattern down to 1 -- 2 minutes is not common, and frequency drifts of tens of MHz/s seem to be unexpectedly large for AD Leo with a rotation period of 2.23 days (compared with frequency drifts of a few MHz/s for stars with a rotation period of a few hours \citep{2015ApJ...802..106L,2022ApJ...935...99B}). To produce such a beaming pattern, we might need to introduce magnetic field lines with very special geometry and certain viewpoint.

Despite this, it is also possible that the frequency drifts are associated with the moving trigger of the radio emission. Adopting a frequency drift of 20 MHz/s and a magnetic field gradient of 0.0036 G/km, the frequency drift corresponds to a source motion at a speed of $2\times10^3$ km/s. If the electron accelerations are due to Alfvén wave trains, the velocity may stand for the local Alfvén speed. The upward and downward frequency drifts might be related to Alfvén waves propagating downwards at first and then getting reflected near the stellar surface.

\subsection{Physical origin}\label{sec:phyori}

\subsubsection{Does the radio emission truly come from AD Leo?}

The previous discussion in this paper is anchored on the belief that the radio emission originates from the target star, AD Leo. However, some may suspect that the radio signals could result from RFIs or other nearby celestial bodies. 

RFIs usually exert a universal influence on all the beams. As a result, we should expect a commonality of the dynamic spectra for different beams if the signals are RFIs. We have checked the data from all the other beams and found no similar signals in the data. Therefore, the possibility of the RFIs can be excluded.

The beams of the FAST receiver have a half-power beamwidth of $\sim$ 3 arcmin at L-band \citep{2020RAA....20...64J}. It means that the objects within an angular distance of 1.5 arcmin will contribute at least half of the radio emission. We noticed that there is a bright background radio source located $\sim$ 2.3' away from AD Leo \citep{1995A&A...295..123S} which will leak a portion of radio emission to the central beam. But from the experience of the previous observations \citep{2006ApJ...637.1016O,2008ApJ...674.1078O}, we could reasonably assume that the background source will not account for the abrupt radio variations in our observations. Furthermore, if the emission came from the background source, it would also leak to other nearby beams and leave traceable signals in the data. We believe that the radio bursts in our cases are generally in accord with the typical characteristics (intensity, polarization, frequency drift) of stellar radio bursts and should be attributed to a radio active star. We could not fully exclude the origin of an unknown radio active star in the vicinity of AD Leo, but the chance should be very low in view of numerous radio events reported on AD Leo.

There is another hypothesis that the radio emission may originate from an exoplanet in AD Leo's extrasolar system. We consider it hardly possible as the emission in 1000 -- 1500 MHz requires a magnetic field strength of 357 -- 536 G. Such magnetic field is not expected to rise from a planetary dynamo \citep{2011Icar..213...12D}, even for hot Jupiters with a remarkably high magnetic field strength \citep{2017ApJ...849L..12Y,2019NatAs...3.1128C}.

\subsubsection{Does the radio emission originate from a magnetospheric process?}

As AD Leo has a dipole magnetic field which can dominate for a very long time, it is possible that a magnetospheric system exists and the radio emission is driven by a magnetospheric process. Two mechanisms worth discussing are breakdown of co-rotation and SPI. We adopted a broadband time-averaged intensity of 10 mJy and assumed a beaming solid angle of 1.6 sr in conformity with Jovian auroral (non-Io) DAM radio emission \citep{2004JGRA..109.9S15Z}, leading to a radio luminosity of $4\times10^{13}\rm erg\; s^{-1}\; Hz^{-1}$. This magnitude of power is typical for ultracool dwarfs with rotation periods less than $\sim 3$ h \citep{2012ApJ...760...59N} while the rotation of AD Leo will result in a luminosity at least two orders of magnitude weaker (see Figure 9 in \cite{2011MNRAS.414.2125N}). Therefore, co-rotation breakdown can not generate emission as intense as we observed. For SPI, we adopted a beaming solid angle of 0.16 sr in accord with Jovian Io-DAM radio emission \citep{2000JGR...10516053K,2001P&SS...49..365Q}, which results in a radio luminosity of $4\times10^{12}\rm erg\; s^{-1}\; Hz^{-1}$. Though the radio power due to SPI is heavily dependent on the parameters of the stellar wind and the potential planet, and many of these are poorly constrained, the predicted SPI power for many M-dwarfs may reach the estimated magnitude \citep{2021NatAs...5.1233C}.

Previous radio observations have failed to detect any long-term periodic signals from AD Leo, though there should be if the emission comes from a magnetospheric process. We suspect that it might be related to incomplete phase coverage for a stellar rotation period or a planetary orbital period in observations. With further radio monitoring, the periodic pattern may prove evident in the future. In addition, it is believed that the auroral emission should have a relatively long duration (several hours) and display some overall frequency drifts, namely the beaming pattern, while the events in our observations are much shorter in time. One possibility is that our results are strongly biased by an incomplete emission pattern. As there might be radio emission beyond the current bandpass, a wider frequency coverage is necessary to have a full view of the integral structures. It is also possible that the supposed long-duration emission is somehow fragmented and we only collect some separated pieces of the emission structures in the observations. As a whole, the SPI is an open possibility for the radio emission, but its exact role should be validated by further investigations.

\subsubsection{Does the radio emission originate from stellar flares in the corona?}

Generally speaking, solar radio bursts are regarded as a by-product of the solar energy release process. Moreover, many nearby M-dwarfs with frequent and intense magnetic activities are also considered among the radio active stars in some recent radio surveys\citep{2019ApJ...871..214V,2021NatAs...5.1233C}. Therefore, it is easy to speculate a connection between the radio emission and stellar flares. Flares powered by magnetic reconnection could accelerate electrons and produce gyrosynchrotron emission due to electron gyromotion. However, it is not fully understood how solar and stellar flares could give rise to coherent ECM radio bursts. A possible mechanism is that the energetic electrons are confined in a closed magnetic loop and gradually form the loss-cone distribution which is responsible for the ECM instability. In our present observations, we fail to detect any coincident optical flares, while there is a flare signal 2 -- 3 hours prior to the radio events on Dec. 3rd. It is worth noting that \cite{2020ApJ...905...23Z} reported a long-duration frequency-drifting radio burst $\sim$ 3 hours after an optical flare and identified it as a flare-induced type IV radio burst. If the radio bursts in our observations are related to the preceding optical flare, how to explain the time delay? A possible explanation is that the emission source is not in an observable location when the flare occurs. As the ECM emission is sharply beamed, we could only detect the coherent emission when the emitter is rotated to a desired longitude, leading to time inconsistency.

There is another possibility that the radio bursts are caused by flares that are not detected in our optical observations. It is suggested that the corona of AD Leo may constantly experience small-scale flares \citep{2003ApJ...582..423G}, some of which might not show up in the optical light curves. As solar flares are more easily detected in extreme-ultraviolet and X-ray bands and the white-light flares only take up a very small fraction \citep{2018ApJ...867..159S}, it is reasonable to speculate that the number of stellar flares is extremely underrated in the observation time period. If the small-scale, transient energy release incidents are the real trigger, they may account for the short duration and random occurrence of the observed radio events.

\section{Discussion}\label{sec:dis}

In Section \ref{sec:obs1} and Section \ref{sec:obs2}, we elaborate on the observational results on Dec. 2nd and Dec. 3rd, respectively. The main results are summarized as follows.

Dec. 2nd

(1) The radio bursts span for $\sim$ 8 minutes and cover the whole frequency band of 1000 -- 1500MHz.

(2) The radio bursts are structured in the form of numerous sub-bursts which display a stripe-like shape and a uniform drift to higher frequencies.

(3) Radio sub-bursts are quasi-periodic in a certain time range. The typical frequency is $\sim$ 5 Hz.

(4) The radio emission is right-hand circularly polarized.

(5) Distribution in the parameter space shows that the drift rate and degree of circular polarization are related to the frequency of the radio sub-burst.

(6) Distribution in the time-frequency plane shows two inverted V-shape patterns. The typical overall frequency drift rate is 19 -- 36 MHz/s.

Dec. 3rd

(1) The radio bursts span for $\sim$ 1.5 hours and are mainly detected below 1150 MHz.

(2) Radio sub-bursts display a blob-like shape with slight elongation. Most of them show a negative frequency drift.

(3) Some of the radio sub-bursts are lined up and form the radio sub-burst trains while others seem to occur randomly.

(4) In certain cases, radio sub-bursts tend to gather in pairs or clusters.

(5) In one event, many radio sub-bursts occur simultaneously at approximately equally-spaced frequencies in the time-frequency plane.

(6) The radio emission is right-hand circularly polarized.

(7) The isolated radio events have a timescale from seconds to minutes. The longest event lasts for $\sim$ 18 minutes and reveals many stripe-like structures.

(8) Multi-wavelength observations reveal no simultaneous detection of flare activities in optical and radio bands.

Quantitative information of the radio sub-bursts in the two days of observations has been listed in Table \ref{tab:table1}. We discuss the possibilities of the two coherent emission mechanisms and explain that the ECM mechanism is the most likely one in Section \ref{sec:emimec}. We offer some possible interpretations on the observed phenomena in Section \ref{sec:pheno} and discuss some mainstream views of stellar radio emission in Section \ref{sec:phyori}. We believe that the radio emission is either the planet-induced SPI signals or the flare-induced radio bursts.

The SPI theories suppose that the sub-Alfvénic interaction between the stellar magnetosphere and an orbiting planet can give rise to the Alfvén waves which propagate along the planet-associated flux tube. The Alfvén waves can efficiently accelerate the electrons which then produce the radio emission in the polar regions of the star. The detected radio emission on AD Leo may imply a potential exoplanet that needs to be confirmed by follow-up exoplanet surveys. Planet-induced radio aurora also indicates the recurrence of radio emission which can be tested with the growing reports of radio observations on AD Leo. A competitive scenario is the flare-induced emission. Considering the flaring nature of AD Leo, magnetic reconnections may instead play the dominant role in producing energetic electrons. The accelerated electrons in the stellar flares can gradually get trapped in the magnetic field lines in the active regions and drive the intense ECM emission. They may also leak to the global-scale field lines and join the magnetospheric currents, but such interactions are not clear right now. We also acknowledge the possibility of a mixed scenario if an M-dwarf is both magnetically-active and planet-hosting. Some radio events may arise from the SPI while some others are related to stellar flares. As a comparison, Jovian DAM radiation has different origins. In simple terms, the emission can be distinguished as Io-DAM and non-Io DAM based on its association with the moon Io using the central meridian longitude (CML)-Io phase diagram \citep{2017A&A...604A..17M}. If the emission is embedded in a global magnetosphere, its time-frequency pattern can be predicted using the well-established simulation in Jupiter studies \citep{2019A&A...627A..30L}. Relevant work will be presented in a follow-up paper.

Despite the macroscopic scenario, fine structures in the ECM emission may reveal important details in the production of ECM emission, which are also worth attention. We found two morphological types of fine structures, which are respectively similar to Jovian S-bursts and solar radio spikes. Currently, there have been several competing theories to explain the generation of ECM fine structures, including Alfvén wave modulation discussed before, electron/ion phase space holes \citep{2006A&ARv..13..229T}, "micro-traps" \citep{1998PhyU...41.1157F}, and so forth. Almost all of them are subject to some related theoretical problems and could only account for parts of the observational results. Though interesting fine structures are revealed in our observations, their underlying mechanisms are largely unclear in the current status. We do not know if the differences in the two days of observations are related to entirely different processes, or the same process but under different plasma conditions. But we are optimistic that the fine structures in the stellar radio bursts will expand the application scope of the ECM emission theories and may contribute to the interpretation of the physical origins in the near future.

\section{Summary}\label{sec:sum}

In this research, we present the high-quality and intriguing results from the first high time-resolution observation of AD Leo with FAST. We detected many radio bursts in the two rounds of observations and managed to resolve the fine structures in the dynamic spectra. We found that each event is composed of numerous millisecond-scale radio sub-bursts which display different morphology on the two days. Radio sub-bursts on Dec. 2nd display a stripe-like shape and a uniform drift to higher frequencies, while the sub-bursts on Dec. 3rd show a blob-like shape with slight elongation and a mostly negative frequency drift. All the events show right-hand circular polarization. We conducted a statistical analysis of the detected radio bursts and discussed their distributions in the parameter space. We generally conclude that the radio bursts are likely ECM emission and that SPI or stellar flares are responsible for the emission.

Our work has fully demonstrated the unique advantage of FAST in stellar radio astronomy. FAST could resolve radio signals from stars with a short sampling time, which is superior in studying the time evolution or even fine structures of the emission. We believe that long-term target observations are necessary to fully understand the physical origin of the emission from a star. With occurrence phase analysis, we could see if the radio events are modulated by the rotation of the star and orbiting of a potential planet. Multi-wavelength coordinated observations may also help to clarify some radio transients. Moreover, the Commensal Radio Astronomy FAST Survey \citep[CRAFTS]{2018IMMag..19..112L} can observe a broad, unbiased sample of stars and detect radio transients in a blind search. A comprehensive radio study can therefore be conducted to see how the radio emission varies from stars of different types. Meanwhile, we look forward to the recommissioning of the ultra-wideband receiver \citep{2019SCPMA..6259502J} which has a wider bandpass (0.27-1.62 GHz) and may cover more topics in extrasolar space weather studies, including radio detection of CMEs and aurora in exoplanets. 

%% IMPORTANT! The old "\acknowledgment" command has be depreciated. It was
%% not robust enough to handle our new dual anonymous review requirements and
%% thus been replaced with the acknowledgment environment. If you try to 
%% compile with \acknowledgment you will get an error print to the screen
%% and in the compiled pdf.
%% 
%% Also note that the akcnowlodgment environment does not support long amounts of text. If you have a lot of people and institutions to acknowledge, do not use this command. Instead, create a new \section{Acknowledgments}.
\begin{acknowledgments}
We are grateful for the financial support from NSFC grants 12250006 \& 11825301. PZ acknowledges funding from the ERC under the European Union’s Horizon 2020 research and innovation programme (grant  agreement no. 101020459 - Exoradio). CKL’s work at the Dublin Institute for Advanced Studies was funded by the Science Foundation Ireland Grant 18/FRL/6199. This work made use of the data from FAST (Five-hundred-meter Aperture Spherical radio Telescope). FAST is a Chinese national mega-science facility, operated by National Astronomical Observatories, Chinese Academy of Sciences. We acknowledge the support of the staff of the Xinglong 85 cm telescope. This work was partially supported by the Open Project Program of the CAS Key Laboratory of Optical Astronomy, National Astronomical Observatories, Chinese Academy of Sciences. We acknowledge the support of the staff of the Weihai 1 m telescope. We thank Lei Qian, Yang Gao, Xingyao Chen for their advice in the application of the FAST observation project. We thank Julien Morin for providing information on the magnetic topology of AD Leo from recent ZDI measurements.
\end{acknowledgments}

%% To help institutions obtain information on the effectiveness of their 
%% telescopes the AAS Journals has created a group of keywords for telescope 
%% facilities.
%
%% Following the acknowledgments section, use the following syntax and the
%% \facility{} or \facilities{} macros to list the keywords of facilities used 
%% in the research for the paper.  Each keyword is check against the master 
%% list during copy editing.  Individual instruments can be provided in 
%% parentheses, after the keyword, but they are not verified.

\vspace{5mm}
\facilities{FAST, Xinglong 85 cm telescope, Weihai 1 m telescope}

%% Similar to \facility{}, there is the optional \software command to allow 
%% authors a place to specify which programs were used during the creation of 
%% the manuscript. Authors should list each code and include either a
%% citation or url to the code inside ()s when available.

\software{Astropy \citep{2013A&A...558A..33A,2018AJ....156..123A}, Matplotlib \citep{Hunter:2007}, NumPy \citep{harris2020array}, 
SciPy \citep{2020SciPy-NMeth}, IRAF \citep{1999ASPC..189...35D}}

%% Appendix material should be preceded with a single \appendix command.
%% There should be a \section command for each appendix. Mark appendix
%% subsections with the same markup you use in the main body of the paper.

%% Each Appendix (indicated with \section) will be lettered A, B, C, etc.
%% The equation counter will reset when it encounters the \appendix
%% command and will number appendix equations (A1), (A2), etc. The
%% Figure and Table counter will not reset.

\appendix

\section{Polarization and flux calibration}\label{sec:pol}

In the full-polarization observations, the four polarization paths from the cross-products of the X and Y signals are simultaneously recorded as XX, YY, CR, CI, where XX=$\langle e_xe_x^*\rangle$, YY=$\langle e_ye_y^*\rangle$, CR=$\langle {\rm Re}[e_xe_y^*]\rangle$, CI=$\langle {\rm Im}[e_xe_y^*]\rangle$. Following the stated convention, the Stokes parameters are defined as $I$=XX+YY, $Q$=XX-YY, $U$=2$\times$CR, $V$=$-2\times$CI \citep{2010PASA...27..104V}. We adopted the standardized polarization calibration scheme \citep{2000ApJ...532.1240B,2004ApJS..152..129V,2021RAA....21..282S} which uses Mueller matrix to characterize the instrumental response and polarization leakage. The mapping between the real Stokes parameters $(I, Q, U, V)$ and observed ones $(I', Q', U', V')$ is

\begin{equation}
    \left(\begin{array}{l}
        I'\\ Q'\\ U'\\V'
    \end{array}\right)=G^2
    \left(\begin{array}{llll}
        \cosh(2\gamma) &\sinh(2\gamma) &0 &0\\ 
        \sinh(2\gamma) &\cosh(2\gamma) &0 &0\\ 
        0 &0 &\cos(2\varphi) &\sin(2\varphi)\\
        0 &0 &-\sin(2\varphi) &\cos(2\varphi)
    \end{array}\right)
    \left(\begin{array}{l}
        I\\Q\\U\\V
    \end{array}\right)
\end{equation}

 where $G$ is the absolute gain, $\gamma$ is the differential gain, and $\varphi$ is the differential phase. We assumed the noise diode signals as 100\% linearly polarized signals, meaning $I=U$, $Q=V=0$. Then, we could derive the differential gain and phase from the full-Stokes noise diode spectra. $\gamma$ is generally between -0.2 and 0.2 and $\varphi$ ranges from $-90^{\circ}$ to $90^{\circ}$ in different frequency channels.

We then referred to the test report of the noise diode on the 19-Beam Receiver in Jan. 2022 \footnote{https://fast.bao.ac.cn/cms/article/150/} to determine the frequency-dependent temperature of the reference signal. We compared the reported noise temperature and the Stokes \emph{I} intensity of the signal to obtain the conversion coefficient between the digital output and antenna temperature.

The follow-up observations of 3C286 were used to calibrate the absolute gain of the telescope, which is the conversion coefficient between the antenna temperature (K) and the flux density (Jy). We referred to \cite{2017ApJS..230....7P} for the polynomial expression of the flux density of 3C286. The total antenna temperature in the central beam during the 3C286 observations includes the contribution from 3C286 and system temperature, the latter one was derived from formula (7) in \cite{2020RAA....20...64J} with the zenith angle equal to $30^{\circ}$. The system temperature is also frequency-dependent and ranges at 24 -- 33 K. We obtained the gain around 15 K/Jy, which yields an aperture efficiency of $\sim$ 59\% compared to the ideal gain (25.6 K/Jy). The derived aperture efficiency is close to the reference value in \cite{2020RAA....20...64J} for the zenith angle of 30$^{\circ}$. However, we could not calibrate the parallactic angle as we do not have the central beam off-point observation on the same day and could not determine the background component of Stokes \emph{Q}, \emph{U}, \emph{V}.

\section{Radio sub-burst auto detection method}\label{sec:pro}

We developed a method to automatically detect radio sub-bursts in the dynamic spectra. The methodology is described in the steps below.

(1) We generated the dynamic spectra by applying the procedures outlined in Section \ref{sec:fdr}, including noise subtraction, polarization and flux calibration, RFI flagging, and background removal. The dynamic spectra of Dec. 2nd were produced with a time (frequency) resolution of 1.6 ms (0.49 MHz) and those of Dec. 3rd have a time (frequency) resolution of 0.8 ms (0.49 MHz).

(2) We used a 2$\times$2 kernel Gaussian filter on the two-dimensional dynamic spectra to reduce background noise fluctuations while preserving main features.

(3) We employed a peak detection algorithm based on persistent topology\footnote{https://www.sthu.org/code/codesnippets/imagepers.html} \citep{10.1007/978-3-658-32182-6_13} to identify the local peaks in the filtered images. The algorithm searched for the peaks in the images by relative amplitude and stopped when the local maximum falls below 5 times the noise level. 

(4) We used a region-growing algorithm to determine the radio sub-bursts' regions in the filtered images, starting from the peaks' locations in step 3 and tracing nearby pixels until their value fell below 3 times the noise level. Pixels in previously defined regions were flagged to avoid overlap.

(5) We calculated the sub-burst properties using the regions defined in step 4. Central frequency and peak time correspond to the peak's location in the time-frequency domain. To calculate the frequency drift rate, we found the pixels with a maximum value at different time steps and performed a linear fit of their locations. The intensity was defined as the average value of the pixels in the region from the raw dynamic spectrum. Time (frequency) width was derived from the number of pixels in the region at the central frequency (peak time). After examining a few sub-burst events, the results automatically calculated using the method closely align with the results that we obtained manually.

\begin{figure*}
	\centering
	\includegraphics[width=0.9\linewidth]{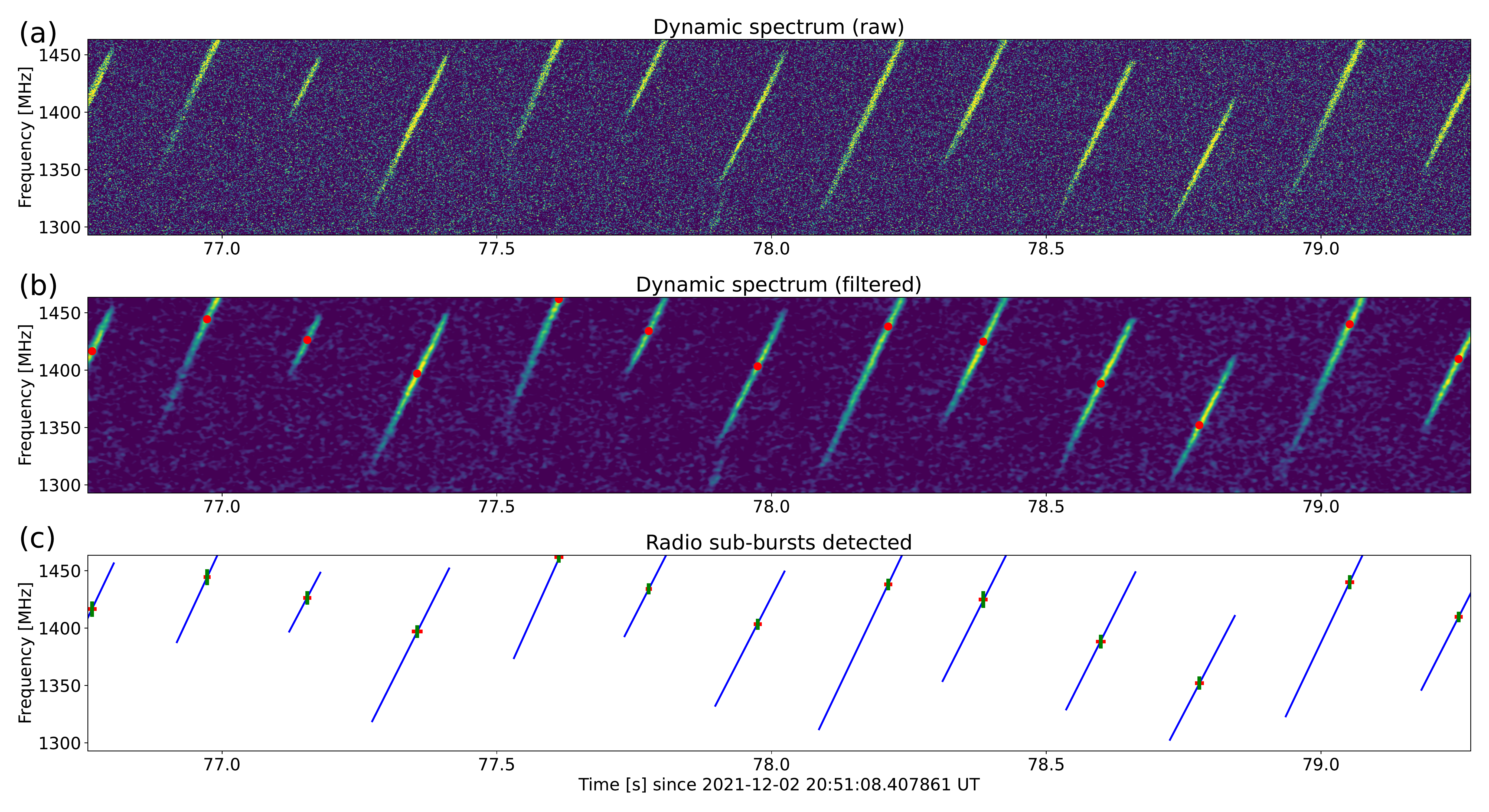}
	\caption{Performance of the method to identify radio sub-bursts on Dec. 2nd. (a) Raw dynamic spectrum. (b) Dynamic spectrum after the Gaussian filter. The red dots stand for the locations of the detected local peaks. (c) Simplified morphology of the detected radio sub-bursts. The blue lines indicate the frequency drift of the sub-bursts. The red and green bars represent the derived time width and frequency width, respectively. \label{fig:figure11}}
\end{figure*}

\begin{figure*}
	\centering
	\includegraphics[width=0.9\linewidth]{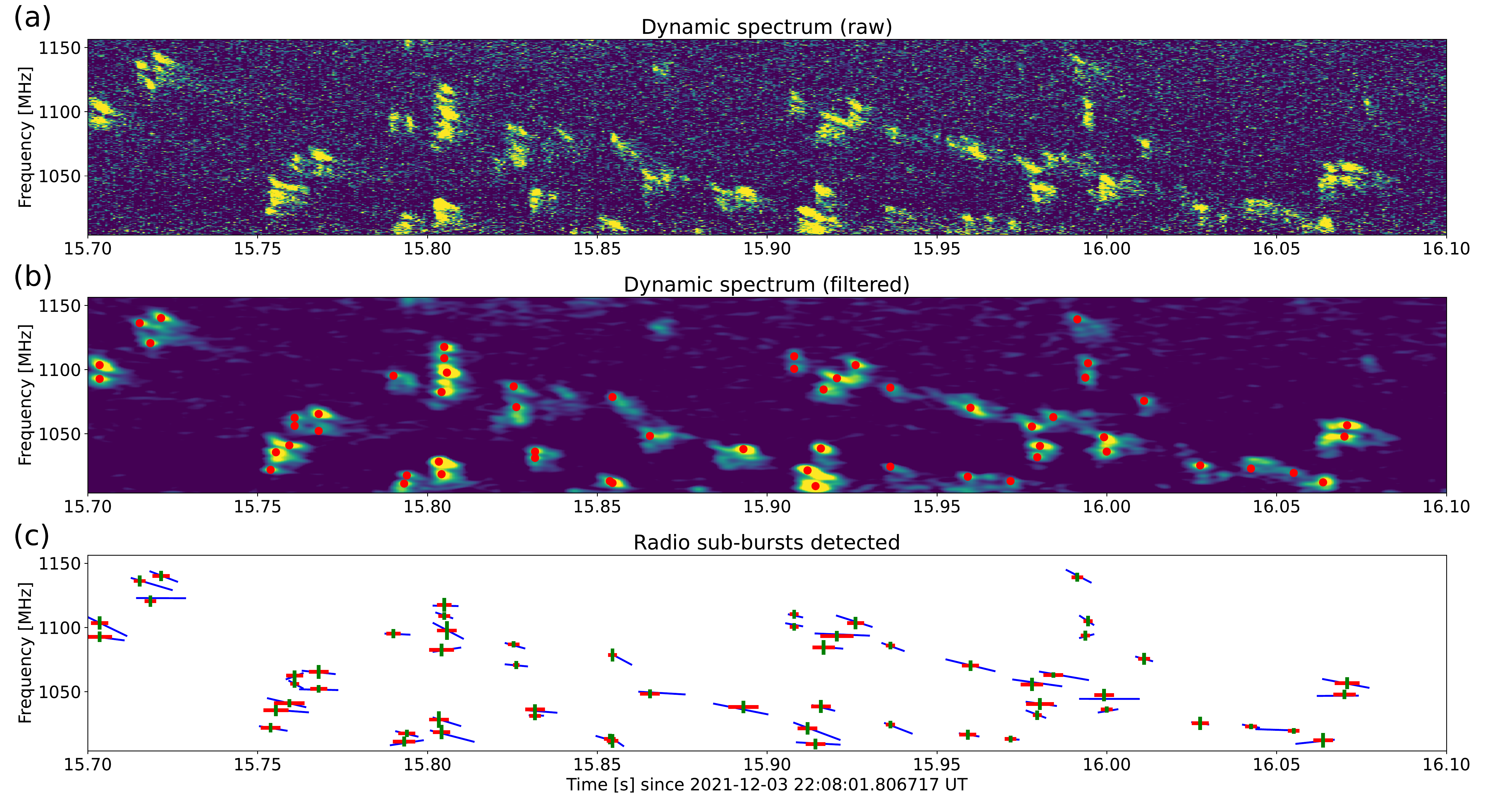}
	\caption{Performance of the method to identify radio sub-bursts on Dec. 3rd. The layouts are the same as in Figure \ref{fig:figure11}.\label{fig:figure12}}
\end{figure*}

The performances of the method to detect radio bursts on Dec. 2nd and Dec. 3rd are shown in Figure \ref{fig:figure11} and Figure \ref{fig:figure12}, respectively.

%% For this sample we use BibTeX plus aasjournals.bst to generate the
%% the bibliography. The sample631.bib file was populated from ADS. To
%% get the citations to show in the compiled file do the following:
%%
%% pdflatex sample631.tex
%% bibtext sample631
%% pdflatex sample631.tex
%% pdflatex sample631.tex

\bibliography{PASPsample631}{}
\bibliographystyle{aasjournal}

%% This command is needed to show the entire author+affiliation list when
%% the collaboration and author truncation commands are used.  It has to
%% go at the end of the manuscript.
%\allauthors

%% Include this line if you are using the \added, \replaced, \deleted
%% commands to see a summary list of all changes at the end of the article.
%\listofchanges

\end{document}